\documentclass[twocolumn,showpacs,prb,citeautoscript]{revtex4}
\usepackage{graphicx}
\usepackage{dcolumn}
\usepackage{bm}
\usepackage{amsmath}
\usepackage{amssymb}

\begin{document}

\title{Microscopic nonequilibrium theory of double-barrier Josephson junctions}
\author{A.\ Brinkman}
\author{A.A.\ Golubov}
\author{H.\ Rogalla}
\affiliation{Faculty of Science and Technology and MESA+ Research
Institute,\\University of Twente, 7500 AE, Enschede, The
Netherlands}
\author{F.K.\ Wilhelm}
\affiliation{Sektion Physik and CeNS,
Ludwig-Maximilians-Universit\"at,\\ Theresienstr. 37, D-80333
M\"unchen, Germany}
\author{M.Yu.\ Kupriyanov}
 \affiliation{Institute of Nuclear Physics, Moscow State University, 119899 Moscow, Russia}

\date{\today }

\begin{abstract}
We study nonequilibrium charge transport in a double-barrier
Josephson junction, including nonstationary phenomena, using the
time-dependent quasiclassical Keldysh Green's function formalism.
We supplement the kinetic equations by appropriate time-dependent
boundary conditions and solve the time-dependent problem in a
number of regimes. From the solutions, current-voltage
characteristics are derived. It is understood why the
quasiparticle current can show excess current as well as deficit
current and how the subgap conductance behaves as function of
junction parameters. A time-dependent nonequilibrium contribution
to the distribution function is found to cause a non-zero averaged
supercurrent even in the presence of an applied voltage. Energy
relaxation due to inelastic scattering in the interlayer has a
prominent role in determining the transport properties of
double-barrier junctions. Actual inelastic scattering parameters
are derived from experiments. It is shown as an application of the
microscopic model, how the nature of the intrinsic shunt in
double-barrier junctions can be explained in terms of energy
relaxation and the opening of Andreev channels.
\end{abstract}
 \maketitle

\section{Introduction}

The Josephson effect is a hallmark of superconductivity. It is
also the basis of a wide range of applications such as metrology,
sensing and classical and quantum logic circuits. Moreover, the
detailed study of the Josephson effect in mesoscopic devices
provides deep insight in the mechanism of the formation and
transport of superconducting correlations in weak links and at
interfaces. It is known from the microscopic theory of
superconductivity that the supercurrent across weak links is
carried by Andreev bound states (ABS). The supercurrent depends
both on the ABS energy levels as well as on their population, i.e.
on the quasiparticle distribution function over energy. This
provides a possibility for external control of current. It was
realized long time ago that deviations from equilibrium may
strongly modify the transport properties of weak links and
Josephson tunnel junctions. A review of early work on various
aspects of nonequilibrium superconductivity is given by the
articles in Refs. ~\onlinecite{Larkin} and ~\onlinecite{Gray} and
by Kopnin \cite{Kopnin}. Two main topics in this field are the
effects arising from a charge imbalance \cite{Tinkham2,Clarke} and
effects from a stimulation of superconductivity by external fields
\cite{Eliashberg}.

The recent progress in the fabrication of superconducting
structures of sub-micrometer size has stimulated a renewed
interest in nonequilibrium effects in Josephson junctions. The
effect of supercurrent control by current injection from
additional terminals was first studied theoretically
\cite{Volkov,Wilhelm,Yip} and demonstrated experimentally
\cite{Morpurgo,Baselmans} for a diffusive Superconductor-Normal
metal-Superconductor (SNS) junction. In this case, even a sign
reversal of the critical current is possible
\cite{Baselmans,Wilhelm}. Additionally, control of supercurrent by
current injection was studied in structures with ballistic
transport \cite{Wees,
Samuelsson,Ilhan2,Thomas,Richter,Samuelsson2}.

Deviations from equilibrium are also reflected in the
quasiparticle current. The dissipative current component in SNS
junctions arises from Multiple Andreev Reflections (MAR) of
quasiparticles. A quasiparticle gains $eV$ in energy each time it
traverses the interlayer, resulting in a strong nonequilibrium
distribution function at subgap energies, as observed by Pierre
\textit{et al} \cite{Pierre}. In point contacts, the microcscopic
description of the current in terms of MAR was derived by Averin
and Bardas \cite{Bardas}, based on a scattering matrix approach,
while Cuevas \textit{et al.} \cite{Cuevas} described MAR in
quantum point contacts by means of a tunnel Hamiltonian approach.
For the case of an SNS junction with a long interlayer as compared
to the coherence length (incoherent regime), the current due to
MAR was calculated by Bezuglyi \textit{et al} \cite{Bezuglyi2}.

When tunnel barriers (I) are introduced at the SN interfaces, the
quasiparticles in short junctions can undergo transmission
resonances, resulting in a dephasing of the electrons and holes.
However, it was shown in Ref. ~\onlinecite{Brinkman} that for a
broad transmission resonance in a SINIS junction, the resonance
energy width being larger than the ABS, coherent transport occurs,
and a microscopic model was given in terms of MAR, integrated by a
universal distribution of transparency eigenvalues \cite{Schep}.
It is shown by Naveh \textit{et al.} \cite{Naveh} that the same
distribution function also describes electrical transport in high
critical current density junctions. In order to model the
nonstationary and nonequilibrium transport through SINIS
structures in the general case, a full Keldysh Green's function
approach is required. The derivation of this microscopic model as
well as its solutions is the scope of this article.

SINIS junctions are promising basic elements for applications in
classical computing and metrology, because they are intrinsically
shunted. Both the Rapid Single Flux Quantum logic \cite{RSFQ} and
the digital voltage standards are electronic applications for
which the use of these structures seems promising
\cite{Kupriyanov2}. However, important features of the $IV$
characterstics are not sufficiently understood yet, such as the
magnitude of the subgap conductance and the nature of the
intrinsic shunt of these junctions. The experimental observation
that the hysteresis in the $IV$ curves depends nonmonotonically on
critical current density has not yet been explained
\cite{Kupriyanov2,Kupriyanov3}. Thus, understanding the transport
properties on a microscopic level will be valuable for electronic
applications. Here, these transport phenomena will be clarified in
terms of the opening of Andreev channels and inelastic scattering
in the interlayer.

Earlier work concentrated on modeling the $IV$ characteristics of
double-barrier junctions in specific limiting cases. When one of
the electrodes is replaced by a normal metal, the
time-dependencies simplify considerably, since formally one can
then put voltage to zero in the superconductor. The $IV$
characteristics of SININ junctions were studied by means of the
quasiclassical Green's functions technique by Zaitsev
\cite{Zaitsev90}, Volkov \textit{et al.} \cite{Volkov2}, and
Zaitsev \textit{et al} \cite{Zaitsev99}. Lempitskii
\cite{Lempitskii} studied nonequilibrium effects on the
nonstationary properties of long SNS junctions in the absence of
interface barriers and Kadin \cite{Kadin} used a time-dependent
Ginzburg-Landau approach, valid only in a narrow temperature
range. Another limiting case is the double-barrier structure with
a long interlayer as compared to the coherence length. The
derivation of time-dependent transport properties in this case
simplifies since a decoupling of the electrodes is possible, as
e.g. studied by Volkov and Klapwijk\cite{Volkov3} and Bezuglyi
\textit{et al} \cite{Bezuglyi2}.

In this article, a microscopic quasiclassical theory will be given
for double-barrier Josephson junction with two superconducting
electrodes and a short interlayer. The interlayer will be assumed
to be a diffusive normal metal, but it will be indicated how the
model can be extended in a straightforward way to incorporate a
superconducting gap in the interlayer. The Keldysh formalism is
introduced in section \ref{section:Keldysh}. The spectral
supercurrent density is obtained and appropriate time-dependent
boundary conditions are derived to supplement the kinetic
equations for the energy distribution functions in the interlayer.
The technical scheme for solving the time-dependent Keldysh-Usadel
equation may have applications beyond the present paper. Solutions
are presented in section \ref{section:adiabatics} for the
adiabatic limit of $eV \ll \Delta_S$. As an intriguing
nonequilibrium effect in a double-barrier Josephson junction, we
show that even at finite voltage bias, there can be a nonzero
averaged supercurrent. Energy relaxation due to inelastic
scattering is a phenomenon that strongly modifies the energy
distribution function. It will be shown in section
\ref{section:scattering} that inelastic scattering is important in
a double-barrier Josephson junction and how this effect can be
incorporated in the microscopic model. As an application of the
microscopic model, the nature of the intrinsic shunt of
double-barrier junctions will be discussed in section
\ref{section:application}. The observed nonmonotonic hysteresis
vs. critical current density dependence, as well as the actual
values, are explained.

\section{Keldysh formulation} \label{section:Keldysh}

The Matsubara Green's function technique can be applied to a
many-body system in equilibrium, from which the energy-dependent
properties of the system can be derived. In addition to obtaining
spectral quantities, we need to know how the states are populated
under nonequilibrium conditions. For this purpose Keldysh
\cite{Keldysh} proposed a set of propagators along a contour in
the complex-time plane that allows to describe the real-time
evolution of a system outside equilibrium and at a finite
temperature. The review of Rammer and Smith \cite{Rammer}
describes the use of the Keldysh technique in the transport theory
of metals. The Keldysh method is introduced specifically for
nonequilibrium superconductivity in Refs.
~\onlinecite{Schmid,Schon,Larkin2,Belzig}.

The quasiclassical approximation is used, in the sense that rapid
oscillations of the wavefunctions on the scale of the
Fermi-wavelengt are averaged out. Furthermore, in this paper it is
assumed that the transport through the interlayer is diffusive,
the thickness being much larger than the elastic scattering
length, so that the Usadel equation can be used.

\subsection{Time-dependent Usadel equation}

A compact notation of the equations for the quasiclassical Green's
functions becomes possible by introducing the Green's function in
Keldysh $\times$ Nambu space
\begin{equation}
\mathord{\buildrel{\lower3pt\hbox{$\scriptscriptstyle\smile$}}
\over G}  = \left( {\begin{array}{*{20}c}
   {\hat G^R } & {\hat G^K }  \\
   0 & {\hat G^A }  \label{Keldysh}
\end{array}} \right).
\end{equation}
The quasiclassical Green's function
$\mathord{\buildrel{\lower3pt\hbox{$\scriptscriptstyle\smile$}}
\over G}$ is a function of two times, $t$ and $t'$, and the
time-dependent Usadel equation in the absence of a
vector potential reads \cite{Larkin2}
\begin{eqnarray}
  - \mathcal D\hbar \nabla \left( {\mathord{\buildrel{\lower3pt\hbox{$\scriptscriptstyle\smile$}}
\over G}  \circ \nabla
\mathord{\buildrel{\lower3pt\hbox{$\scriptscriptstyle\smile$}}
\over G} } \right) &+&
\mathord{\buildrel{\lower3pt\hbox{$\scriptscriptstyle\smile$}}
\over \tau } _3 \hbar \frac{{\partial
\mathord{\buildrel{\lower3pt\hbox{$\scriptscriptstyle\smile$}}
\over G} }}{{\partial t}} + \frac{{\partial
\mathord{\buildrel{\lower3pt\hbox{$\scriptscriptstyle\smile$}}
\over G} }}{{\partial t'}}\hbar
\mathord{\buildrel{\lower3pt\hbox{$\scriptscriptstyle\smile$}}
\over \tau } _3  -
i\mathord{\buildrel{\lower3pt\hbox{$\scriptscriptstyle\smile$}}
\over \Delta } \left( t
\right)\mathord{\buildrel{\lower3pt\hbox{$\scriptscriptstyle\smile$}}
\over G} \nonumber \\ +
\mathord{\buildrel{\lower3pt\hbox{$\scriptscriptstyle\smile$}}
\over G}
i\mathord{\buildrel{\lower3pt\hbox{$\scriptscriptstyle\smile$}}
\over \Delta } \left( {t'} \right)
  &=&  - i\left( {\mathord{\buildrel{\lower3pt\hbox{$\scriptscriptstyle\smile$}}
\over \Sigma_{\rm inel} }  \circ
\mathord{\buildrel{\lower3pt\hbox{$\scriptscriptstyle\smile$}}
\over G}  -
\mathord{\buildrel{\lower3pt\hbox{$\scriptscriptstyle\smile$}}
\over G}  \circ
\mathord{\buildrel{\lower3pt\hbox{$\scriptscriptstyle\smile$}}
\over \Sigma_{\rm inel} } } \right),\label{Usadel}
\end{eqnarray}
where
\begin{eqnarray}
\mathord{\buildrel{\lower3pt\hbox{$\scriptscriptstyle\smile$}}
\over \tau } _3  = \left( {\begin{array}{*{20}c}
   {\hat \tau _3 } & 0  \\
   0 & {\hat \tau _3 }  \\
\end{array}} \right),{\rm{ }}\mathord{\buildrel{\lower3pt\hbox{$\scriptscriptstyle\smile$}}
\over \Delta }  = \left( {\begin{array}{*{20}c}
   {\hat \Delta } & 0  \\
   0 & {\hat \Delta }  \\
\end{array}} \right), \nonumber \\{\rm{ }}\hat \Delta  = \left( {\begin{array}{*{20}c}
   0 & \Delta   \\
   {\Delta ^* } & 0  \\
\end{array}} \right),{\rm{ }}\mathord{\buildrel{\lower3pt\hbox{$\scriptscriptstyle\smile$}}
\over \Sigma_{\rm inel} }  = \left( {\begin{array}{*{20}c}
   {\hat \Sigma_{\rm inel} } & 0  \\
   0 & {\hat \Sigma_{\rm inel} }  \\
\end{array}} \right),
\end{eqnarray}
$\mathcal D$ is the diffusion constant, $
\mathord{\buildrel{\lower3pt\hbox{$\scriptscriptstyle\smile$}}
\over \Sigma_{\rm inel} } $ the self energy with retarded,
advanced and Keldysh components, * denotes the complex conjugate
and $\circ$ denotes a convolution over the internal time
coordinates, e.g.
$\mathord{\buildrel{\lower3pt\hbox{$\scriptscriptstyle\smile$}}
\over \Sigma_{\rm inel} }  (t,t^\prime) \circ
\mathord{\buildrel{\lower3pt\hbox{$\scriptscriptstyle\smile$}}
\over G}  = \int {dt_1
\mathord{\buildrel{\lower3pt\hbox{$\scriptscriptstyle\smile$}}
\over \Sigma_{\rm inel} } \left( {t,t_1 }
\right)\mathord{\buildrel{\lower3pt\hbox{$\scriptscriptstyle\smile$}}
\over G} } \left( {t_1 ,t'} \right) $. The function
$\mathord{\buildrel{\lower3pt\hbox{$\scriptscriptstyle\smile$}}
\over G} $ is normalized as
$\mathord{\buildrel{\lower3pt\hbox{$\scriptscriptstyle\smile$}}
\over G}  \circ
\mathord{\buildrel{\lower3pt\hbox{$\scriptscriptstyle\smile$}}
\over G}  =
\mathord{\buildrel{\lower3pt\hbox{$\scriptscriptstyle\smile$}}
\over 1} $. The expression for the current in the Keldysh
formalism is
\begin{equation}
I = \frac{1}{{2eR_N }}\int {dE{\rm{Tr}}} \left[ {\hat \tau _3
\left( {\hat G^R \nabla \hat G^K  + \hat G^K \nabla \hat G^A }
\right)} \right]. \label{current}
\end{equation}

The Green's functions can be transformed to energy-frequency space
$\left({E,\omega} \right)$ by Fourier transforming the functions
$\mathord{\buildrel{\lower3pt\hbox{$\scriptscriptstyle\smile$}}
\over G} \left( {t - t',\left( {t + t'} \right)/2} \right)$,
\begin{eqnarray}
\mathord{\buildrel{\lower3pt\hbox{$\scriptscriptstyle\smile$}}
\over G} \left( {E,\omega } \right) &=& \int
{\mathord{\buildrel{\lower3pt\hbox{$\scriptscriptstyle\smile$}}
\over G}} \left( {t - t',\frac{{t + t'}}{2}} \right)e^{ - iE(t -
t')/\hbar } \nonumber
\\&& e^{i\omega (t + t')/2\hbar } d\left( {t
- t'} \right)d\left( {t + t'} \right)/2 .
\end{eqnarray}
This transformation is analogous to the Wigner representation of
the full double-{\em coordinate} Green's function. Spectral
quantities that only depend on energy and not on frequency after
Fourier transforming, such as the equilibrium Green's functions in
the electrodes, only depend on the time difference before Fourier
transforming. Each term in Eq.~(\ref{Usadel}) can be transformed
to $\left({E,\omega} \right)$-space. Hence, the Usadel equation
can be rewritten in $\left({E,\omega} \right)$-space as
\begin{eqnarray}
 &&- \mathcal D\hbar \nabla \left( {\mathord{\buildrel{\lower3pt\hbox{$\scriptscriptstyle\smile$}}
\over G}  \circ \nabla
\mathord{\buildrel{\lower3pt\hbox{$\scriptscriptstyle\smile$}}
\over G} } \right) + iE\left[
{\mathord{\buildrel{\lower3pt\hbox{$\scriptscriptstyle\smile$}}
\over \tau } _3 , \circ
\mathord{\buildrel{\lower3pt\hbox{$\scriptscriptstyle\smile$}}
\over G} } \right] + i\frac{\omega }{2}\left\{
{\mathord{\buildrel{\lower3pt\hbox{$\scriptscriptstyle\smile$}}
\over \tau } _3 , \circ
\mathord{\buildrel{\lower3pt\hbox{$\scriptscriptstyle\smile$}}
\over G} } \right\} \nonumber \\&&=  - i\left(
{\mathord{\buildrel{\lower3pt\hbox{$\scriptscriptstyle\smile$}}
\over \Sigma_{\rm inel} }  \circ
\mathord{\buildrel{\lower3pt\hbox{$\scriptscriptstyle\smile$}}
\over G}  -
\mathord{\buildrel{\lower3pt\hbox{$\scriptscriptstyle\smile$}}
\over G}  \circ
\mathord{\buildrel{\lower3pt\hbox{$\scriptscriptstyle\smile$}}
\over \Sigma_{\rm inel} } } \right), \label{Usadel2}
\end{eqnarray}
where $\hat \Delta \left(t \right) = 0$ is taken for simplicity. $
[ {\mathord{\buildrel{\lower3pt\hbox{$\scriptscriptstyle\smile$}}
\over \tau } _3
,\mathord{\buildrel{\lower3pt\hbox{$\scriptscriptstyle\smile$}}
\over G} }] $ is the commutator of $
\mathord{\buildrel{\lower3pt\hbox{$\scriptscriptstyle\smile$}}
\over \tau } _3 $ and
$\mathord{\buildrel{\lower3pt\hbox{$\scriptscriptstyle\smile$}}
\over G} $, and $ \{
{\mathord{\buildrel{\lower3pt\hbox{$\scriptscriptstyle\smile$}}
\over \tau } _3
,\mathord{\buildrel{\lower3pt\hbox{$\scriptscriptstyle\smile$}}
\over G} } \} $ is the anti-commutator. A decomposition of the
Green's functions in Fourier harmonics can formerly be introduced
as
\begin{equation}
\mathord{\buildrel{\lower3pt\hbox{$\scriptscriptstyle\smile$}}
\over G} \left( {E,\omega } \right) = \sum\limits_{n =  - \infty
}^\infty
{\mathord{\buildrel{\lower3pt\hbox{$\scriptscriptstyle\smile$}}
\over G} _n \left( E \right)\delta \left( {\omega  - \omega _0 }
\right)} , \end{equation} where
$\mathord{\buildrel{\lower3pt\hbox{$\scriptscriptstyle\smile$}}
\over G} _n \left( E \right) =
\mathord{\buildrel{\lower3pt\hbox{$\scriptscriptstyle\smile$}}
\over G} \left( {E,n\omega _0 } \right)$, as was for example done
in Ref. \onlinecite{Cuevas}. The generation of higher order
Fourier harmonics is a manifestation of the nonlinearity of the
device, prevalent e.g.\ in Eq.\ (\ref{Usadel2}).

\subsection{Retarded and advanced propagators}

Equation (\ref{Usadel2}) consists of an Usadel equation for the
retarded Green's function, the advanced Green's function, and an
equation containing the Keldysh Green's function. The Usadel
equation for the retarded Green's function $ \hat G^R$ in the
interlayer (taking the limit of $\hat \Delta = 0$ and zero
inelastic scattering, $\hat \Sigma_{\rm inel} =0$) in Fourier components
reads
\begin{eqnarray}
 &&- \mathcal D\hbar \nabla \left( {\hat G^R  \circ \nabla \hat G^R } \right)_n  + in\omega _0 /2\left\{ {\hat \tau _3 ,\hat G_n^R \left( E \right)} \right\} \nonumber
 \\&&+ iE\left[ {\hat \tau _3 ,\hat G_n^R \left( E \right)} \right] =
 0,\label{Fourier}
\end{eqnarray}
where the index $n$ denotes the $n$-th harmonic. The self-energy
terms in Eq.~(\ref{Usadel2}) can effectively be represented by a
characteristic inelastic scattering time $\tau _{in} $, as was
derived by Larkin and Ovchinnikov \cite{Larkin3}. Taking the
inelastic scattering to be time-independent is of course a rough
approximation, but suffices for the purposes of this paper. The
self-energy terms have been neglected in Eq.~(\ref{Fourier}),
which is justified as long as $\hbar /\tau _{in}  \ll k_B T$.
Note, that Eq.\ (\ref{Fourier}) can be reduced to the
time-independent case for $n=0$. Time-dependence occurs if the
boundary conditions, which will be detailed below, provide nonzero
limiting values for $\hat{G}_n$.

\begin{figure}
\includegraphics [scale=1.25]{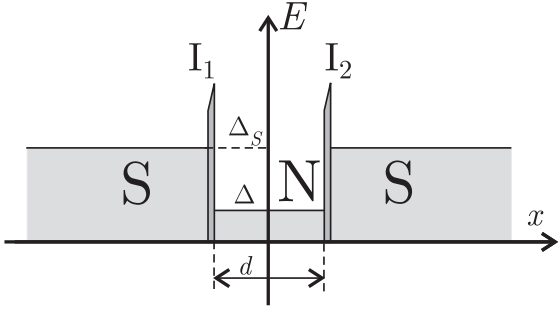}
\caption{\label{fig0} Schematic representation of the
double-barrier SINIS structure. Two superconducting electrodes (S)
are separated by two delta-shaped potential barriers (I$_1$ and
I$_2$) and a normal metal interlayer (N). The position dependence
of the pair potential has been indicated by shading.}
\end{figure}

The most general decomposition of the retarded and advanced
Green's functions is a linear combination of the three Pauli
matrices \cite{Rammer}. Whenever the phase is constant, it can be
chosen such that it suffices to define
\begin{eqnarray}
 \hat G^{R(A)}  = G^{R(A)} \hat \tau _3  + F^{R(A)} \hat \tau _1 , \nonumber\\
 \hat G^A  =  - \hat \tau _3 \hat G^{R\dag } \hat \tau _3  =  - G^{R * } \hat \tau _3  + F^{R * } \hat \tau _1
 .\label{def}
 \end{eqnarray}
Assuming that the thickness of the interlayer $d$ is much smaller
than the coherence length in the interlayer, $\xi=\sqrt{{\cal
D}/E}$ where $E$ is a characteristic energy at which the system is
probed, we can take the retarded Green's function much larger than
its gradient. The double-barrier structure under consideration is
depicted in Fig. \ref{fig0}. Integrating both sides of
Eq.~(\ref{Fourier}) over the interlayer thickness and barriers
gives

\begin{eqnarray}
 \hbar \mathcal D \left. {\left( {\hat G^R  \circ \nabla \hat G^R } \right)_n } \right|_ {x = 0^+}  - \hbar \mathcal D\left. {\left( {\hat G^R  \circ \nabla \hat G^R } \right)_n } \right|_ {x = d^-} \nonumber \\
  + in\frac{{\omega _0 }}{2}d\left\{ {\hat \tau _3 ,\hat G_n^R \left( E \right)} \right\} + iEd\left[ {\hat \tau _3 ,\hat G_n^R \left( E \right)} \right] = 0.
 \end{eqnarray}

Zaitsev \cite{Zaitsev84} derived effective boundary conditions for
the quasiclassical Green's function formalism. These were further
developed for diffusive scattering in the interlayer by Kupriyanov
and Lukichev \cite{Kupriyanov}. Using the Kupriyanov-Lukichev
boundary conditions for the retarded Green's functions,
\begin{equation}
 \left. {\xi \gamma _B \left( {\hat
G^R \circ \nabla \hat G^R } \right)_n} \right|_{ x = 0,d} = \pm
\left[ {\hat G^R , \circ \hat G_{S_{L,R} }^R } \right]_n ,
\end{equation}
where $\gamma_B = R_B/\rho \xi$, $\rho$ is the resistivity of the
interlayer, and $R_B$ is the interface resistance, we obtain
\begin{eqnarray}
 &&in\frac{{\omega _0 }}{2}\left\{ {\hat \tau _3 ,\hat G_n^R \left( E \right)} \right\}\gamma _B d/\xi + iE\left[ {\hat \tau _3 ,\hat G_n^R \left( E \right)} \right]\gamma _B d/\xi \nonumber\\
 &&+2\pi k_B T_{cS} \left[ {\hat G^R , \circ \left( {\hat G_{S_R }^R  + \hat G_{S_L }^R } \right)} \right]_n =0,\label{Fourier2}
 \end{eqnarray}
where $\hat G_{S_{L,R}}^R $ are retarded functions in the left and
right electrodes respectively. The normalization condition for $
\hat G^R$ in energy-space and decomposed into Fourier harmonics
can be found from the expressions of Appendix
\ref{appendix:convolutions},
\begin{equation}
\delta _{n0}  = \sum\limits_{m =  - \infty }^\infty  {\hat G_m^R
\left( {E + \frac{{n - m}}{2}\omega _0 } \right)} \hat G_{n - m}^R
\left( {E - \frac{m}{2}\omega _0 } \right).\label{norm}
\end{equation}
Equations (\ref{Fourier2}) and (\ref{norm}) form a complete set of
equations from which the Fourier components of $\hat G^R$ can in
principle be determined. This recursive schemes reflects the fact,
that superconducting correlations can be induced over several MAR
cycles. Solving the set of equations is complicated by the
recurrent nature of the equations. The Fourier harmonics are
coupled to each other and have arguments that are shifted in
energy.

From the full set of equations, we can find back the
quasi-stationary Matsubara case by keeping only the $n = 0$
harmonic of $G^R$ and the $n = \pm 1$ harmonics of $F^R$ and
neglecting energy-shifts in the arguments. This provides a
solution for $G^R$ and $F^R$ that coincides with the analytical
continuation ($\omega \to - iE$) of the Matsubara solution at
$\varphi  = \pi /2$. The general Matsubara solutions for
double-barrier junctions with $\gamma_{B1,2} \gg 1$ and $d/\xi \ll
1 $ were obtained in Ref. ~\onlinecite{Kupriyanov2}. In the limit
of $\Delta = 0$ the analytical continuation $\omega  \to - iE$ of
this solution provides the Green's functions at $T=0$ as function
of energy
\begin{eqnarray}
\Phi &=&\frac{EF_S}{E \gamma _{\rm{eff}}/\pi k_BT_c+iG_S}\left(
\cos \frac \varphi 2+i\gamma _{\_}\sin \frac \varphi 2\right)\nonumber,\\
G&=&\frac E{( E^2-\left| \Phi \right| ^2) },\text{ }F=\frac \Phi
{( \left| \Phi \right| ^2-E^2) }\label{GF},
\end{eqnarray}
where $F_S=\Delta _S/\left( \Delta _S^2-E^2\right)$ and
$G_S=E/\left( E^2-\Delta _S^2\right)$. The asymmetry and effective
suppression parameter are respectively
\begin{eqnarray}
\gamma _ -   = \frac{{\gamma _{B1}  - \gamma _{B2} }}{{\gamma
_{B1}  + \gamma _{B2} }},{\rm{ }}\gamma _{\rm{eff}}  =
\frac{d}{{\xi  }}\frac{{\gamma _{B1} \gamma _{B2} }}{{\gamma _{B1}
+ \gamma _{B2} }}.
\end{eqnarray}

\begin{figure}
\includegraphics [scale=1.25]{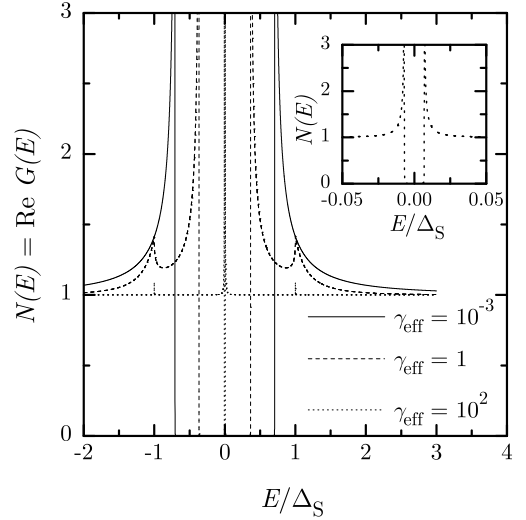}
\caption{\label{fig_dos} Normalized density of states in the
interlayer at $T=0$ for several values of the suppression
parameter $\gamma_{\mathop{\rm eff}\nolimits}$. The inset shows
the minigap that is present for $\gamma_{\mathop{\rm
eff}\nolimits} = 10^2$ on a smaller scale.}
\end{figure}

As an illustration to these retarded Green's functions, the
density of states in the interlayer, $N={\mathop{\rm
Re}\nolimits}G$, is shown in Fig.~\ref{fig_dos}. It can be seen
that for $\gamma_{\mathop{\rm eff}\nolimits} \gg 1$ the density of
states is determined by a minigap with a value of $\cos (\varphi
/2)\pi k_BT_c/\gamma _{\rm {eff}}$. In the coherent regime
\cite{Brinkman} of $\gamma_{\mathop{\rm eff}\nolimits} \ll 1$ the
gap in the density of states is given by $\Delta \cos (\varphi
/2)$. The density of states for intermediate values of the
suppression parameter is characterized by a two-peak structure.
These findings coincide with the calculations of Bezuglyi
\textit{et al.} \cite{Bezuglyi} in the limiting case of a short
interlayer.

\subsection{Spectral supercurrent}

Supercurrent is carried by states in the weak link and their
occupation is determined by a distribution function. The
supercurrent-carrying density of states, or spectral supercurrent
${\mathop{\rm Im}\nolimits} I_S \left( E \right)$, can be
determined from $\hat G^R$ and $\hat G^A$ by
\begin{equation}
{\mathop{\rm Im}\nolimits} I_S  = \frac{1}{8}{\rm{Tr}}\left[ {\hat
\tau _3 \left( {\hat G^R  \circ \nabla \hat G^R  - \hat G^A  \circ
\nabla \hat G^A } \right)} \right].
\end{equation}

\begin{figure}
\includegraphics [scale=1.25]{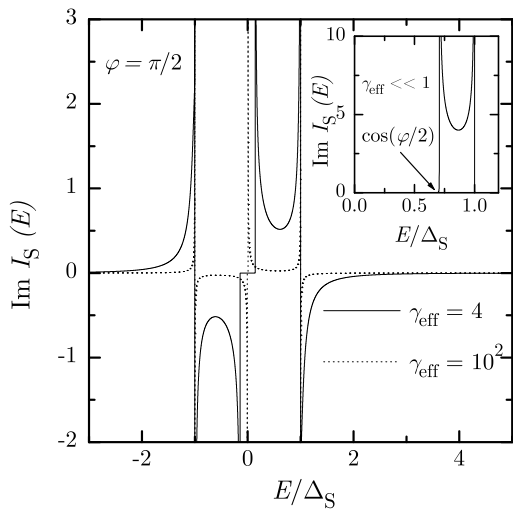}
\caption{\label{fig_sc} Normalized spectral supercurrent density
as a function of energy for various values of the suppression
parameter $\gamma_{\mathop{\rm eff}\nolimits}$. The phase
difference between the superconducting electrodes was fixed at
$\varphi = \pi /2$. The inset shows the spectral supercurrent in
the coherent regime of $\gamma_{\mathop{\rm eff}\nolimits} \ll
1$.}
\end{figure}

The supercurrent in the regime of $\gamma_{\mathop{\rm
eff}\nolimits} \ll 1$, is found to have a spectral density
\begin{equation}
{\mathop{\rm Im}\nolimits} I_S \left( E \right)eR_N  =
\frac{{\Delta _S^2 \sin \varphi }}{{\sqrt {\Delta _S^2  - E^2 }
\sqrt {E^2  - \Delta _S^2 \cos ^2 \left( {\varphi /2} \right)} }},
\end{equation}
for $\Delta _S \cos ( {\varphi /2} ) < E < \Delta _S $, while $
{\mathop{\rm Im}\nolimits} I_S ( E ) = 0$ for $E < \Delta _S \cos
( {\varphi /2} )$ and $E > \Delta _S $. This universal expression
is independent of the interlayer thickness, barrier height and
contact dimensionality as long as the number of conduction
channels is large \cite{Galaktionov}. The same expression was
found in the case of ballistic interlayer transport
\cite{Brinkman}. The spectral supercurrent is nonzero only in the
range $\Delta _S \cos ( {\varphi /2} ) < E < \Delta _S $, i.e.
there is a minigap $\Delta _S \cos ( {\varphi /2} )$ in the
spectrum of the Andreev bound states, see the inset of
Fig.~\ref{fig_sc}. On the other hand, all states in the energy
range $\Delta _S \cos ( {\varphi /2} )$ contribute to the
supercurrent. In long junctions, similar behavior for the DOS was
found in Ref.~ \onlinecite{Charlat} and for the current in Ref.~
\onlinecite{Wilhelm}. The contact is in the intermediate regime
between a short ballistic SNS weak link, with bound state energy
$\Delta _S \cos ( {\varphi /2} )$ and a tunnel junction with bound
state energy $\Delta _S $. Physically this is caused by the
properties of the distribution of transparencies, which is a
combination of open and closed channels (see Ref.
~\onlinecite{Brinkman}).

In the incoherent regime of $\gamma_{\mathop{\rm eff}\nolimits}
\gg 1$, this universality breaks down. The minigap in the spectrum
of Andreev bound states is now given by $\cos ( {\varphi /2} ) \pi
k_B T_c / \gamma_{{\mathop{\rm eff}\nolimits}}$. Figure
\ref{fig_sc} shows the spectral supercurrent density for several
values of the suppression parameter. The sign change at
$E=\Delta_S$ has also been observed by Bezuglyi \textit{et al.}
\cite{Bezuglyi} and Heikkil\"a \textit{et al} \cite{Heikkila}.
Going beyond the approximations $d \ll \xi$ and $\gamma_B \ll 1$,
it was shown by Sch\"apers \textit{et al.} \cite{Thomas2} for
ballistic junctions that low-energy states are gradually filled in
for larger interlayer thickness and by larger barrier
transparency.

\subsection{Kinetic equations}\label{section:kinetic}

The energy distribution functions, that determine the occupation
of spectral funtions, can be determined from the kinetic
equations.

From the matrix normalization condition
$\mathord{\buildrel{\lower3pt\hbox{$\scriptscriptstyle\smile$}}
\over G}  \circ
\mathord{\buildrel{\lower3pt\hbox{$\scriptscriptstyle\smile$}}
\over G}  =
\mathord{\buildrel{\lower3pt\hbox{$\scriptscriptstyle\smile$}}
\over 1} $, the upper right component implies that $\hat G^R \circ
\hat G^K  + \hat G^K  \circ \hat G^A  = 0$. Hence, $\hat G^K $ can
be parametrized as
\begin{equation}
\hat G^K  = \hat G^R  \circ \hat f - \hat f \circ \hat G^A.
\label{param}
\end{equation}
Furthermore, it was shown by Schmid and Sch\"on \cite{Schmid2} and
Larkin and Ovchinnikov \cite{Larkin3}, that $\hat f$ can be chosen
to be diagonal. We will adopt the notation
\begin{equation}
\hat f = f_L \hat 1 + f_T \hat \tau _3 ,\label{notation}
\end{equation} where $f_L$ and $f_T$ are those parts of the
distribution function that are respectively even and odd in
energy. Therefore they are named longitudinal and transverse
energy distribution function respectively. The functions can be
identified with energy and particle flow \cite{Peltier}.
Physically, a deviation of $f_L$ from equilibrium is associated
with a different effective temperature and a deviation of $f_T$
from equilibrium with a chemical potential shift. In equilibrium,
$f_{T0} = 0$ and $f_{L0} = \tanh(E/2k_BT)$.

Putting Eqs.~(\ref{param}) and (\ref{notation}) into the Keldysh
component of the Usadel Eq.~(\ref{Usadel}) and by making use of
the Usadel equations for the retarded and advanced Green's
function, finally the kinetic equations for the Fourier components
of $f_L$ and $f_T$ can be written as
\begin{eqnarray}
 &&\left( {D_L  \circ \nabla ^2 f_L } \right)_n  + \left( {{\mathop{\rm Im}\nolimits} I_S  \circ \nabla f_T } \right)_n =\frac{1}{{\hbar \mathcal D}}\left( {ni\omega _0  + \hbar \tau _{in}^{ - 1} } \right)\nonumber \\
 &&\times \left[ {G^R  \circ \left( {f_L  - f_0 \delta _{n0} } \right) - \left( {f_L  - f_0 \delta _{n0} } \right)G^A } \right]_n
 ,\nonumber
 \end{eqnarray}
\begin{eqnarray}
&&\left( {D_T  \circ \nabla ^2 f_T } \right)_n  + \left(
{{\mathop{\rm Im}\nolimits} I_S  \circ \nabla f_L } \right)_n  =
\frac{1}{{\hbar \mathcal D}}\left( {ni\omega _0  + \hbar \tau
_{in}^{ - 1}
} \right) \nonumber\\
&&\times \left( {G^R  \circ f_T  - f_T  \circ G^A } \right)_n
,\label{kinetic}
\end{eqnarray}
with the generalized transverse and longitudinal diffusion
coefficients being $4D_T = {\rm{Tr}}( 1 - \hat \tau _3 \hat G^R
\circ \hat \tau _3 \hat G^A ) $, $ 4D_L = {\rm{Tr}}(1 - \hat G^R
\circ \hat G^A )$. With the parametrization of Eq. (\ref{def})
this can be further rewritten as $D_T =
({\rm{Re}}G)^2+({\rm{Re}}F)^2$ and $D_L =
({\rm{Re}}G)^2-({\rm{Im}}F)^2$. In obtaining Eqs. (\ref{kinetic}),
use has been made of the rewriting of the $
\mathord{\buildrel{\lower3pt\hbox{$\scriptscriptstyle\smile$}}
\over \Sigma }  \circ
\mathord{\buildrel{\lower3pt\hbox{$\scriptscriptstyle\smile$}}
\over G}  -
\mathord{\buildrel{\lower3pt\hbox{$\scriptscriptstyle\smile$}}
\over G}  \circ
\mathord{\buildrel{\lower3pt\hbox{$\scriptscriptstyle\smile$}}
\over \Sigma } $ term by Larkin and Ovchinnikov \cite{Larkin3}
into a collision integral with characteristic inelastic scattering
time $\tau _{in} $. $\Delta$ has been assumed to be negligible for
simplicity, but a superconducting gap in the interlayer can be
incorporated in the model in a straightforward way by keeping the
terms in the Usadel Eq. (\ref{Usadel}) that depend on $\Delta$. In
the limit of slow time variations, a Fourier transform over the
time difference provides the known mixed representation of the
kinetic equations \cite{Larkin2}
\begin{eqnarray}
 &&\mathcal DD_L \nabla ^2 f_L  + \mathcal D{\mathop{\rm Im}\nolimits} I_S \nabla f_T  \nonumber \\&&-{\mathop{\rm Re}\nolimits} G\left( {\tau _{in}^{ - 1}  + \frac{d}{{dt}}} \right)\left( {f_L  - f_0 } \right) = 0, \nonumber\\
 &&\mathcal DD_T \nabla ^2 f_T  + \mathcal D{\mathop{\rm Im}\nolimits} I_S \nabla f_L  \nonumber \\&&-{\mathop{\rm Re}\nolimits} G\left( {\tau _{in}^{ - 1}  + \frac{d}{{dt}}} \right)f_T  =
 0,\label{timekin}
 \end{eqnarray}
which follows directly from Eq.~(\ref{kinetic}) for the lowest
Fourier harmonic. The expressions for the supercurrent and
dissipative current components can be derived \cite{Belzig} from
Eq. (\ref{current})
\begin{eqnarray}
 I_S  = \frac{1}{{2eR_N }}\int {dEf_L \left( E \right){\mathop{\rm Im}\nolimits} I_S \left( E \right)} , \\
 I_N  = \frac{1}{{2eR_N }}\int {dED_T \left( E \right)\nabla f_T \left( E
 \right)}.\label{current2}
\end{eqnarray}
What remains to be derived, is a proper set of time-dependent
boundary conditions for the kinetic equations.

\subsection{Time-dependent boundary conditions}

The Kupriyanov-Lukichev boundary conditions \cite{Kupriyanov} for
the quasiclassical Green's functions can in general be written as
\begin{equation}
\gamma _B \xi
\mathord{\buildrel{\lower3pt\hbox{$\scriptscriptstyle\smile$}}
\over G} \circ
\frac{d}{{dx}}\mathord{\buildrel{\lower3pt\hbox{$\scriptscriptstyle\smile$}}
\over G} =
\mathord{\buildrel{\lower3pt\hbox{$\scriptscriptstyle\smile$}}
\over G} \circ
\mathord{\buildrel{\lower3pt\hbox{$\scriptscriptstyle\smile$}}
\over G} _1  -
\mathord{\buildrel{\lower3pt\hbox{$\scriptscriptstyle\smile$}}
\over G} _1  \circ
\mathord{\buildrel{\lower3pt\hbox{$\scriptscriptstyle\smile$}}
\over G} ,\label{KLbound} \end{equation} where $
\mathord{\buildrel{\lower3pt\hbox{$\scriptscriptstyle\smile$}}
\over G} _1 $ and $
\mathord{\buildrel{\lower3pt\hbox{$\scriptscriptstyle\smile$}}
\over G} $ denote the Green's functions at the two sides of the
first interface. From the definition of the Green's functions in
Keldysh space, Eq. (\ref{Keldysh}), a boundary condition can be
written for each matrix element. In Appendix \ref{appendix:bound},
this set of boundary conditions is rewritten into
\begin{eqnarray}
 && \gamma _B \xi \left[ {\left( {1 - \hat G^R \hat G^A } \right)\frac{d}{{dx}}f_{L}  + \left( {\hat \tau _3  - \hat G^R \hat \tau _3 \hat G^A } \right)\frac{d}{{dx}}f_{T} } \right]
 \nonumber\\  &&= \left[ {\hat G^R \left( {\hat G_1^R  - \hat G_1^A } \right) - \left( {\hat G_1^R  - \hat G_1^A } \right)\hat G^A } \right]\left( {f_{L1}  - f_{L} } \right)
  \nonumber \\ &&+ \left[ {\hat G^R \left( {\hat G_1^R \hat \tau _3  - \hat \tau _3 \hat G_1^A } \right) - \left( {\hat G_1^R \hat \tau _3  - \hat \tau _3 \hat G_1^A } \right)\hat G^A } \right]\nonumber\\ &&\times\left( {f_{T1}  - f_{T} }
  \right),\label{boundstart}
\end{eqnarray}
where all the products have to be regarded as time-convolutions
and $f_{L1/T1}$ are the distribution functions in the respective
reservoir.

The Green's functions become time dependent by applying a voltage
over the interface. In the absence of voltage, the Green's
functions in the electrodes only depend on time difference since
equilibrium is assumed. The potential can be introduced in each
electrode by a Gauge transformation \cite{Zaitsev84} of the
Green's function in the electrodes
\begin{equation}
\hat G_1^{R(A)} \left( {t,t'} \right) = \hat S\left( t \right)\hat
G_1^{R(A)} \left( {t - t'} \right)\hat S^\dag  \left( {t'}
\right),
\end{equation}
where $\hat S\left( t \right)$ and $\hat S^\dag  \left( {t'}
\right) $ are given by
\begin{eqnarray} S\left( t \right) = \left(
{\begin{array}{*{20}c}
   {e^{ieVt/\hbar } } & 0  \\
   0 & {e^{ - ieVt/\hbar } }  \\
\end{array}} \right) \nonumber, \\{\rm{ }}S^\dag  \left( {t'} \right) = \left( {\begin{array}{*{20}c}
   {e^{ - ieVt'/\hbar } } & 0  \\
   0 & {e^{ieVt'/\hbar } }  \label{bound}
\end{array}} \right).
\end{eqnarray}
Volkov and Klapwijk \cite{Volkov3} performed a Gauge
transformation of the interlayer Green's functions, which works
only in the the limit $d \gg \xi $ because of the small coupling
between the electrodes in this case which allows to neglect the
interference terms in the interlayer leading to a local
time-dependence.

By performing the Gauge transformation for $ \hat G_1^R$ and $
\hat G_1^A$ and by taking the trace from Eq. (\ref{boundstart}),
one obtains the first boundary condition in time representation.
The second equation is obtained by taking the trace after
multiplying left- and right-hand side of Eq. (\ref{boundstart}) by
$\hat \tau _3 $. This results in
\begin{eqnarray}
 &&D_T \gamma _B \xi \frac{d}{{dx}}f_T  = \left\{ {{\mathop{\rm Re}\nolimits} G_1 {\mathop{\rm Re}\nolimits} G i\sin \left[ {\frac{{eV}}{\hbar }\left( {t - t'} \right)} \right]} \right. \nonumber\\
 &&\left. { + {\mathop{\rm Im}\nolimits} F_1 {\mathop{\rm Re}\nolimits} F \sin \left[ {\frac{{eV}}{\hbar }(t + t')} \right]} \right\}\left( {f_{L0}  - f_L } \right)\nonumber \\
 && - f_T \left\{ {{\mathop{\rm Re}\nolimits} G_1 {\mathop{\rm Re}\nolimits} G \cos \left[ {\frac{{eV}}{\hbar }\left( {t - t'} \right)} \right]} \right. \nonumber\\
 &&\left. { + {\mathop{\rm Re}\nolimits} F_1 {\mathop{\rm Re}\nolimits} F \cos \left[ {\frac{{eV}}{\hbar }(t + t')} \right]}
 \right\},\label{bound1}
\end{eqnarray}
\begin{eqnarray}
  &&D_L \gamma _B \xi \frac{d}{{dx}}f_L  = \left\{ {{\mathop{\rm Re}\nolimits} G_1 {\mathop{\rm Re}\nolimits} G \cos \left[ {\frac{{eV}}{\hbar }\left( {t - t'} \right)} \right]} \right. \nonumber\\
 &&\left. { - {\mathop{\rm Im}\nolimits} F_1 {\mathop{\rm Im}\nolimits} F \cos \left[ {\frac{{eV}}{\hbar }(t + t')} \right]} \right\} \left( {f_{L0}  - f_L } \right)\nonumber\\
 && - f_T \left\{ {{\mathop{\rm Re}\nolimits} G_1 {\mathop{\rm Re}\nolimits} G i\sin \left[ {\frac{{eV}}{\hbar }\left( {t - t'} \right)} \right]} \right. \nonumber\\
 && + \left. {{\mathop{\rm Re}\nolimits} F_1 {\mathop{\rm Im}\nolimits} F \sin \left[ {\frac{{eV}}{\hbar }(t + t')} \right]} \right\}, \label{bound2}
\end{eqnarray}
where all products are time convolutions and use has been made of
the fact that $f_{T1} = 0$, since the electrodes are assumed to be
in internal equilibrium. The energy distribution functions are not
only coupled through the kinetic equations (\ref{kinetic}), but
through the boundary conditions as well.

At the second interface a similar set of boundary conditions can
be derived, which can be obtained from Eqs. (\ref{bound1}) and
(\ref{bound2}) by replacing $G_1$ and $F_1$ by $G_2$ and $F_2$
respectively, and by multiplying the right-hand side of Eqs.
(\ref{bound1}) and (\ref{bound2}) by $-1$.

Note, that $D_L  = ( {{\mathop{\rm Re}\nolimits} G } )^2  - (
{{\mathop{\rm Im}\nolimits} F } )^2  = 0$ for energies smaller
than the minigap in the interlayer. Hence, for energies at which
$D_L = 0$, boundary condition (\ref{bound2}) is replaced by $f_L =
f_{L0} $. This physically means, that the system does not conduct
heat inside the gap and that the distribution in the gap is
controlled by coupling to some external heat bath, e.g.\ through
the substrate, and {\em not} through the superconducting leads.

Each term in the boundary condition contains time convolutions.
With the aid of the expansion of the Green's functions in Fourier
harmonics and the expressions of Appendix
\ref{appendix:convolutions} for the time convolutions of double
and triple products, the convolutions can be worked out for each
term. The left-hand side of Eq. (\ref{bound1}) is for example
\begin{eqnarray}
  &&D_T  \circ \gamma _B \xi \frac{d}{{dx}}f_{T}  = \sum\limits_{n,n'} {\int\limits_{ - \infty }^\infty  {D_{T,n} \left( {E + n'\omega _0 /2} \right)} } \gamma _B \xi
 \nonumber \\  &&\frac{d}{{dx}}f_{T,n'} \left( {E - n\omega _0 /2} \right)e^{iE\left( {t - t'} \right)/\hbar } e^{i\frac{{n + n'}}{{2\hbar }}\omega _0 \left( {t + t'} \right)}
 dE.\nonumber\\
  \label{term1}
\end{eqnarray} The sine and cosine dependencies in the boundary
conditions cause additional voltage shifts as well as coupling to
higher harmonics, which can be seen for example in the term
\begin{eqnarray}
  &&f_L  \circ \left[ {{\mathop{\rm Re}\nolimits} G_S  \circ {\mathop{\rm Re}\nolimits} G \circ i\sin \left( {t - t'} \right)} \right] \cr
  && = \sum\limits_{n,n'} {\int {dEf_{L,n}} \left( {E + n'\omega _0 /2} \right){\mathop{\rm Re}\nolimits} G_{n'} \left( {E - n\omega _0 /2} \right)}  \nonumber \\ &&\times e^{iE'\left( {t - t'} \right)/\hbar } e^{i\frac{{n + n'}}{{2\hbar }}\omega _0 \left( {t + t'}\right)}\nonumber \\
  && \frac{1}{2}\left[ {{\mathop{\rm Re}\nolimits} G_S \left( {E + \frac{{n' - n - 1}}{2}\omega _0 } \right)} \right. \nonumber\\&& + \left.{ {\mathop{\rm Re}\nolimits} G_S \left( {E + \frac{{n' - n + 1}}{2}\omega _0 } \right)} \right].
  \label{term2}
\end{eqnarray}
In principle, the set of kinetic equations (\ref{kinetic})
together with the boundary conditions Eqs. (\ref{bound1}) and
(\ref{bound2}), and expressions for the time convolutions, such as
Eq. (\ref{term1}) and Eq. (\ref{term2}), now provide a complete
set of equations to solve the energy distribution functions as
function of voltage. However, the coupling to higher harmonics and
energy shifts within the functions themselves make solving the
equations cumbersome. In principle, a solution should crossover to
the solutions as found by the MAR approach \cite{Brinkman} in the
limit of $\gamma_{\mathop{\rm eff}\nolimits} \ll 1$. In the next
section an adiabatic approximation will be developed in order to
solve the kinetic equations for $eV \ll \Delta _S $ and a larger
suppression parameter.

\section{Adiabatic dynamics} \label{section:adiabatics}
\subsection{Adiabatic approximation}

In order to simplify the time-dependencies, an adiabatic
approximation can be made. When voltage is small, the phase
oscillates slowly and can even be considered quasi-stationary. In
this case, we only need to keep the time dependence in expressions
that contain the phase, but can neglect all other time
dependencies. Consequently, the time convolutions become simple
products and the energy shifts can be neglected. Therefore, this
approximation is called adiabatic.

A formal derivation of the parameter regime in which the adiabatic
approximation can be used, is based on the time dependence in Eqs.
(\ref{bound1}) and (\ref{bound2}). The quasiparticle current is
determined by the left-hand side of Eq. (\ref{bound1}), namely
$D_T \gamma _B \xi df_{T} /dx$. It will be shown in this section
that the right-hand side of this boundary condition in the
adiabatic limit is equal to the $f_T$ terms in Eq. (\ref{bound1}).
Hence, deviations from the adiabatic approximation in the
quasiparticle current are only to be expected when the terms
proportional to $f_{L}$ in Eq. (\ref{bound1}) are not negligible.
The first of these terms is $f_{L} {\mathop{\rm Re}\nolimits} G_1
{\mathop{\rm Re}\nolimits} G i\sin [ {eV( {t - t'} )/\hbar } ]$,
which can be neglected for $eV \ll \Delta _S $. The second term is
$f_{L} {\mathop{\rm Im}\nolimits} F_1 {\mathop{\rm Re}\nolimits} F
\sin [ {eV( {t + t'} )/\hbar } ]$, which is nonzero only due to
the loop-like construction with Eq. (\ref{bound2}), in which the
terms ${\rm{Im}}F_1$ and ${\rm{Re}}F$ are shifted $eV/2$ in energy
every cycle, making their overlap nonzero after approximately
$\Delta_S/eV$ cycles. For large suppression parameters,
${\mathop{\rm Re}\nolimits} F \sim \gamma _{{\mathop{\rm
eff}\nolimits}}^{ - 1} $. Hence, smallness of this term can now be
formulated as $( {1/\gamma _{{\mathop{\rm eff}\nolimits}} }
)^{\Delta _S /eV} \ll 1$. Combining the conditions, the conclusion
is reached that the adiabatic approximation is valid when $eV \ll
\Delta _S $ and $\gamma _{{\mathop{\rm eff}\nolimits}} \gg 1$.

In this case, the phases $\chi_{1,2}$ can be introduced by the
parameterization
\begin{equation}
\hat G_{1,2}^R  = \left(
{\begin{array}{*{20}c}
   {G_{1,2}^R } & {F_{1,2}^R {\mathop{\rm e}\nolimits} ^{i\chi _{1,2} } }  \\
   {F_{1,2}^R {\mathop{\rm e}\nolimits} ^{ - i\chi _{1,2} } } & { - G_{1,2}^R }  \\
\end{array}} \right).\label{param2}
\end{equation}
No additional Gauge transformations have to be performed in the
boundary conditions. Hence, we can use the parametrization of Eq.
(\ref{param2}) directly in the boundary condition
(\ref{boundstart}). The first boundary condition is then found by
taking the trace of Eq. (\ref{boundstart}). The second is found by
taking the trace after multiplying with $\hat \tau _3$. With Eq.
(\ref{param2}) and after some rewriting this gives
\begin{eqnarray}
 \gamma _{B1,2} \xi  D_T \frac{d}{{dx}}f_{T} \left( {0,d} \right) =  \pm M_{T1,2} \left[ {f_{T} \left( {0,d} \right) \mp f_{T0}} \right], \nonumber\\
 \gamma _{B1,2} \xi  D_L \frac{d}{{dx}}f_{L} \left( {0,d} \right) =  \pm M_{L1,2} \left[ {f_{L} \left( {0,d} \right) - f_{L0}}
 \right],\label{bound3}
 \end{eqnarray}
where
\begin{equation}
  f_{L0,T0}  = \frac{1}{2}\tanh \frac{{E + eV/2}}{{2k_B T}} \pm \frac{1}{2}\tanh \frac{{E - eV/2}}{{2k_B T}},
\end{equation}
are the distribution functions in the leads and $M_{T1,2}  =
{\mathop{\rm Re}\nolimits} G_{S_{L,R}} {\mathop{\rm Re}\nolimits}
G + {\mathop{\rm Re}\nolimits} F_S {\mathop{\rm Re}\nolimits} F
\cos ( {\chi _1  - \chi _2 } )$, $ M_{L1,2}  = {\mathop{\rm
Re}\nolimits} G_{S_{L,R}} {\mathop{\rm Re}\nolimits} G  -
{\mathop{\rm Im}\nolimits} F_S {\mathop{\rm Im}\nolimits} F \cos (
{\chi _1  - \chi _2 } )$, where $\chi _1 - \chi _2  = \varphi /2 =
eVt$ in the case of symmetric barriers. Here, $G$ and $F$ are
given by Eq. (\ref{GF}), and
\begin{equation}
 G_{S_{L,R}}  = \frac{{E \pm
 eV/2}}{{\sqrt {\left( {E \pm eV/2} \right)^2  - \Delta _S^2 }
 }},{\rm{ }}F_S  = \frac{{\Delta _S }}{{\sqrt {\Delta _S^2  - E^2 }
 }}.
\end{equation}
As can be seen from here, using also section
\ref{section:kinetic}, the kinetic equations in the
quasi-stationary limit coincide with the known adiabatic equations
in the mixed representation as derived by Larkin and Ovchinnikov
\cite{Larkin2}. In the absence of inelastic scattering, the
equations further simplify to
\begin{eqnarray}
 D_T \left( E \right)\frac{{d^2 f_T}}{{dx^2 }} \left( {E,x} \right) + {\mathop{\rm Im}\nolimits} I_S \left( E \right)\frac{df_L}{{dx}} \left( {E,x} \right) = 0,
 \nonumber\\
 D_L \left( E \right)\frac{{d^2 f_L}}{{dx^2 }} \left( {E,x} \right) + {\mathop{\rm Im}\nolimits} I_S \left( E \right)\frac{df_T}{{dx}} \left( {E,x} \right) =
 0.\label{kinetic2}
\end{eqnarray}
Therefore, Eqs. (\ref{bound3}) and (\ref{kinetic2}) provide the
set of equations to describe the time-dependent transport in
double-barrier junctions in the limit of a large suppression
parameter and a small voltage.

\subsection{Dissipative current}\label{section:qp}

From the expression for the dissipative current component in Eq.
(\ref{current2}), and by solving Eqs. (\ref{bound3}) and
(\ref{kinetic2}) with an Ansatz $f_{T,L}  = a_{1,2} x^2  + b_{1,2}
x + c_{1,2} $, it is obtained that
\begin{eqnarray}
&&I_{qp} \left( t \right) = \frac{1}{{eR_N }}\int\limits_{ -
\infty }^\infty  {\frac{{dE}}{2}} \frac{{\gamma _{B1}  + \gamma
_{B2} }}{{\frac{{\gamma _{B1} }}{{M_{T1} }} + \frac{{\gamma _{B2}
}}{{M_{T2} }}}} f_{T0}.\label{qp}
\end{eqnarray}
This solution is obtained in the limit of no inelastic scattering.
The effects of energy relaxation in the interlayer will be
discussed in section \ref{section:inelastic}. It follows from Eq.
(\ref{qp}) that the quasiparticle current has in general a
phase-dependent contribution through the coefficients $M_{T1,2}$.
The current is therefore time-dependent since the phase difference
is given by $\varphi  = 2eVt/\hbar $. The dc component is then
determined by averaging over time.

When one of the superconducting electrodes is replaced by a normal
metal, the expressions for $M_{T,L}$ simplify, since $F_N = 0$. By
putting voltage to zero in the superconductor, it can be shown
that the expression for the quasiparticle current, Eq. (\ref{qp})
with $M_{T2} = $Re$G$, coincides with the known results of the
SININ junction of Volkov \textit{et al} \cite{Volkov2}. The
additional term $m\left( E \right) = d^{ - 1} \int {dx/M_T \left(
{E,x} \right)} $ in Ref. ~\onlinecite{Volkov2} is neglected in our
case, since the term is small as compared to $\gamma _B /M_T$.

\begin{figure}
\includegraphics [scale=1.25]{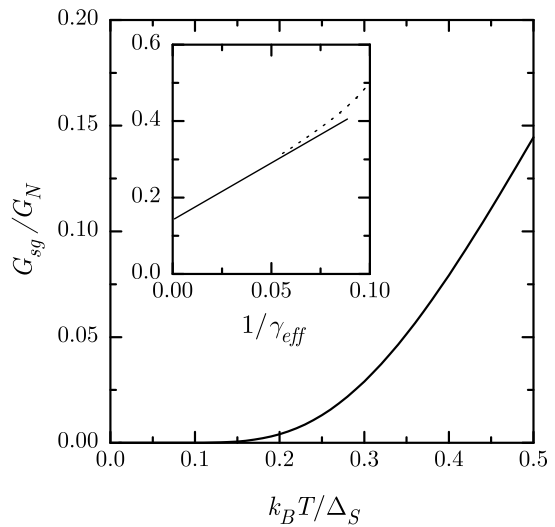}
\caption{\label{fig3_4_a} Normalized conductance at $eV =
\Delta_S$ as function of temperature (normalized to $\Delta_S$)
for $\gamma_{{\mathop{\rm eff}\nolimits}} \gg 1$. The inset shows
the normalized conductance at $eV = \Delta_S$ as function of
$1/\gamma_{{\mathop{\rm eff}\nolimits}}$ at $k_BT/\Delta_S =
0.5$.}
\end{figure}

\begin{figure}
\includegraphics [scale=1.25]{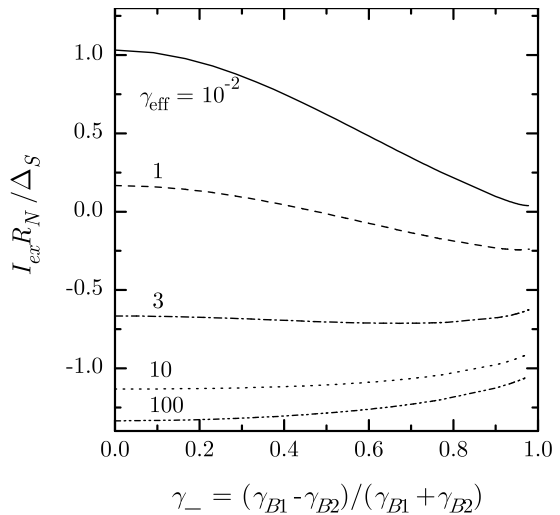}
\caption{\label{fig3_4_b} Excess and deficit current as function
of asymmetry for several values of the suppression parameter.}
\end{figure}

As a measure of the subgap conductance of a double-barrier
Josephson junction, the conductance at $eV = \Delta_S$ is
calculated and shown in Fig.~\ref{fig3_4_a} as function of
temperature in the limit of $\gamma _{{\mathop{\rm eff}\nolimits}}
\gg 1$. The inset of Fig.~\ref{fig3_4_a} shows the conductance at
$eV = \Delta_S$ as function of the inverse suppression parameter.
It can be seen that the conductance is enhanced by a decrease in
$\gamma_{{\mathop{\rm eff}\nolimits}}$. Physically, this
corresponds to the opening of Andreev channels due to the term
${\mathop{\rm Re}\nolimits} F_S {\mathop{\rm Re}\nolimits} F$ in
$M_T$. For $\gamma_{{\mathop{\rm eff}\nolimits}}  > 10$, for which
the model of this section is a good approximation, the conductance
is found to be approximately proportional to $\gamma_{{\mathop{\rm
eff}\nolimits}}^{ - 1} $. This proportionality will be used in
section \ref{section:application} to predict an intrinsic shunt in
high-$J_c$ double-barrier Josephson junctions.

\textit{High voltage bias.} In the regime of a large voltage $eV
\gg \Delta _S $, the time dependencies in the electrodes become
decoupled and the current in an SINIS junction can be seen as the
summation of the current in an SININ' junction and an N'INIS
junction. In this case, the relevant functions $M_T$ in Eq.
(\ref{qp}) simplify, and the presence of excess and deficit
current can be calculated. Figure~\ref{fig3_4_b} shows the
resulting dependence of the excess and deficit current on the
asymmetry parameter, for several values of the suppression
parameter. Note, that the decoupling into an SININ' and N'INIS
junction at $eV \gg \Delta _S $ is valid for all values of
$\gamma_{{\mathop{\rm eff}\nolimits}}$. The limiting case of a
deficit current $eI_{def} R_N  = 4\Delta _S /3$ for the symmetric
limit and $\gamma_{{\mathop{\rm eff}\nolimits}} \gg 1$ coincides
with the findings of Zaitsev \cite{Zaitsev84} and Volkov
\textit{et al} \cite{Volkov2}. The excess current $eI_{ex} R_N
\simeq 1.05\Delta _S $ for $\gamma_{{\mathop{\rm eff}\nolimits}}
\ll 1$ coincides with the results of MAR calculations
\cite{Brinkman}.

\subsection{Nonequilibrium supercurrent at finite voltage}

From Eq. (\ref{current2}) and a solution of the kinetic equations,
the supercurrent can be determined as a function of voltage. In
most tunnel junctions and weak links, the time-dependence of the
spectral supercurrent is harmonic and by averaging over time, the
supercurrent becomes zero at a finite voltage. However, due to the
additional time dependence of $f_L$, the product of $f_L$ and
Im$I_S$ not necessarily has to be harmonic, and a nonzero
time-averaged supercurrent can exist at a finite voltage.
Physically, the time-dependence of $f_L$ originates from the fact
that due to the proximity effect, the heat diffusion coefficient
depends on the phase difference, as is known e.g. from Andreev
interferometers \cite{Peltier}.  We would like to calculate
\begin{equation}
I_S\left( V\right) =\frac 1{2eR_N}\left\langle \int f_L(E,t)
{\mathop{\rm Im}\nolimits} I_S \left( E,t \right) dE \right\rangle
_t,
\end{equation}
where the brackets denote time-averaging. It can be seen that this
expression explicitly depends on the spectral supercurrent, which
could for example be suppressed by a magnetic field. Therefore,
this current contribution is true supercurrent in the presence of
an applied voltage.

Since the time-dependent perturbation of $f_L$ is much smaller
than $f_0$, the kinetic equations (\ref{timekin}) can be rewritten
in the adiabatic form like Eqs. (\ref{kinetic2}), now including
inelastic scattering,
\begin{eqnarray}
D_T \frac{{d^2 f_T}}{{dx^2 }}  + {\mathop{\rm Im}\nolimits} I_S
\frac{df_L}{{dx}}  = \delta^{-1}f_T,
 \nonumber\\
 D_L \frac{{d^2 f_L}}{{dx^2 }} + {\mathop{\rm Im}\nolimits} I_S \frac{df_T}{{dx}}  =
 \delta^{-1}(f_L-f_0),\label{kinetic3}
\end{eqnarray}
where, $\delta$ is introduced as $\delta ^{-1}=N\xi ^2/\mathcal
D\tau _{in}$. Under the conditions of the adiabatic approximation,
we can use again boundary conditions (\ref{bound3}), and with
Ansatz solutions $f_T=a_1x^2+b_1x+c_1$ and $f_L=a_2x^2+b_2x+c_2$,
this provides us with the solution
\begin{equation}
f_L-f_0=\gamma _{\rm{eff}} {\mathop{\rm Im}\nolimits} I_S \frac
M{\gamma _B D_T} f_{T0}\label{fT_time}
\end{equation}
where
\begin{equation}
M=\frac{2M_{T1}M_{T2}+2\gamma _{\rm{eff}}\delta
^{-1}M_{T1}}{\left( M_{T1}+M_{T2}+2\gamma _{\rm{eff}}\delta
^{-1}\right) \left( M_{L1}+M_{L2}+2\gamma _{\rm{eff}}\delta
^{-1}\right) }.\label{M}
\end{equation}

From Eqs. (\ref{fT_time}) and (\ref{M}) it can be seen that in the
limit of strong inelastic scattering, $\tau_{in} \to 0$, $f_L-f_0$
will be proportional to $\tau_{in}$ and therefore equilibrium is
restored.

\begin{figure}
\includegraphics [scale=1.25]{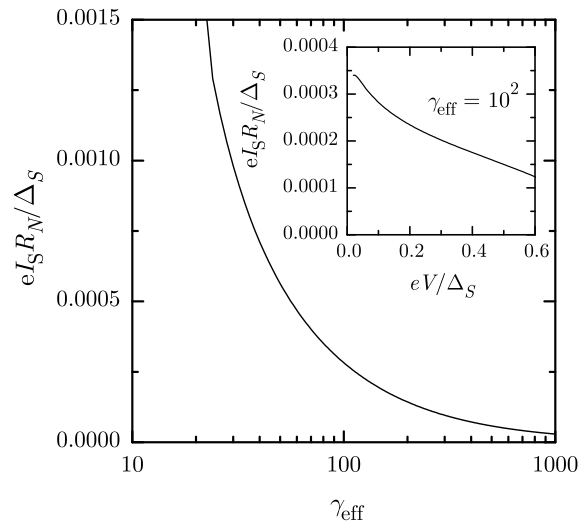}
\caption{\label{fig_Is} Nonequilibrium averaged supercurrent at a
finite voltage bias ($eV=0.1\Delta_S$) at $k_BT=0.2\Delta_S$ and a
small inelastic scattering rate $\gamma=10^{-3}\Delta_S$ as
function of the suppression parameter. The inset shows the
supercurrent as function of bias voltage at $k_BT=0.2\Delta_S$ for
a fixed suppression parameter $\gamma_{\rm{eff}}$.}
\end{figure}

The nonequilibrium correction to $f_L$ is obtained within the
adiabatic approximation. Therefore, we assume that time dependence
only comes into the final expressions via the phase factor in the
spectral supercurrent density Im$I_S$. In the case of symmetric
barriers, ${\rm{Im}}^2I_S$ averaged over time is equal to the
average of sine-squared, which is just a factor 1/2. The
supercurrent contribution can therefore be written as
\begin{equation}
I_S=\frac 1{2eR_N}\gamma _{\rm eff}\int\limits_0^\infty D_T^{-1}\left( {\rm{Im}}%
I_S\right) ^2f_{T0}MdE.
\end{equation}
In order to perform the integral some smearing of the
${\rm{Im}}I_S^2$ divergency has to be assumed. Physical reasons
for this are always present, like a small amount of inelastic
scattering. An inelastic scattering term $\gamma$ can be taken
into account in the retarded part of the Usadel equations
\cite{Rammer}, but in the limit of little inelastic scattering,
$\gamma \ll k_BT_{cS}$, the scattering term can be presented in
the solutions by transforming the energy $E$ to $E+i \gamma$. In
section \ref{section:application}, realistic values of the
inelastic scattering parameter and the suppression parameter will
be derived. From these values it can be concluded that
$\gamma_{{\mathop{\rm eff}\nolimits}} \delta ^{-1} \ll 1$ in most
of the practical cases. In this limit, the only contribution to
the supercurrent comes from $E>\Delta_S$, the spectral
supercurrent being zero below the minigap and $M$ being zero
between the minigap and $\Delta_S$. The resulting supercurrent in
this limit is shown in Fig.~\ref{fig_Is} as function of the
suppression parameter and in the inset as function of voltage.

The physics of this effect is similar to the physics of the
nonequilibrium supercurrent first considered by Lempitskii
\cite{Lempitskii} for a long diffusive SNS junction without
potential barriers at the NS interfaces. The mechanism is the
conversion of quasiparticle current into supercurrent inside the
junction, formally described by the coupling term ${\mathop{\rm
Im}\nolimits} I_S df_T/dx$ in Eq. (\ref{kinetic3}). An alternative
explanation for this mechanism has been given in terms of
thermoelectricity in Ref. ~\onlinecite{Peltier}. However,
quantitative differences occur between the cases of long and short
junctions. In a long SNS junction with $d\gg\xi$ the strongest
deviations of the distribution function $f_L$ from equilibrium
occur at the sub-gap energy range, at energies of the order of the
Thouless energy $\hbar \mathcal D/d^2$. In the case of MAR, an
additional non-equilibrium correction to $f_L$ appears, as shown
by Pierre \textit{et al.} \cite{Pierre}, which is beyond the
present adiabatic approach in which voltage and interface
transparencies are small so that MAR is suppressed. At sub-gap
energies the excitation of the symmetric mode described by $f_L$,
generated by the quasiparticle current, cannot diffuse out of the
junction since the corresponding diffusion constant $D_L$ vanishes
in the S-electrodes. On the other hand, in SINIS junctions, the
symmetric mode at low bias is excited only at $E>\Delta_S$ since
the quasiparticle current vanishes in the energy range
$E<\Delta_S$ due to the presence of tunnel barriers. Hence,
deviations of $f_L$ from equilibrium occur only at $E>\Delta_S$.
Since the diffusion constant in the S-electrodes $D_L<1$ at
$E>\Delta_S$, the excitations at $E>\Delta_S$ are partially
trapped at these energies, though the magnitude of the effect in
SINIS junctions is smaller than in SNS junctions. With a
decreasing barrier heigth in SINIS junctions, an Andreev
contribution to the quasiparticle current appears and deviations
from equilibrium should also occur at $E<\Delta_S$. The study of
this crossover, however, is beyond the scope of this paper.

The nonequilibrium supercurrent at a certain voltage is
approximately two order of magnitude smaller than the dc
supercurrent at zero voltage, but is comparable with the
quasiparticle current at the same voltage. The latter conclusion
is based on the assumption that inelastic scattering is
negligible, but we will see in the next section that energy
relaxation in the interlayer increases the subgap quasiparticle
current. Therefore we can conclude that the nonequilibrium
supercurrent can be present in double-barrier junctions, but that
under realistic junction parameter values, the contribution to the
total current is minor. Note, that the nonequilibrium supercurrent
in the considered regime is proportional to the square of the sine
of the phase difference over the junction, $I_S \sim {\rm sin}^2
\phi$. This effect can give rise to the occurence of half-integer
Shapiro steps in the $IV$ characteristics when the junction is
irradiated by microwaves \cite{Argaman,Dubos}.

\section{Inelastic scattering} \label{section:scattering}

In many mesoscopic systems and weak links, the time of flight of a
quasiparticle through a normal metal or superconducting layer is
much shorter than the characteristic inelastic scattering time in
the specific material. Hence, inelastic scattering in mesoscopic
systems and weak links is usually neglected. However, in
double-barrier junctions, time of flight can be large. Because of
the normal reflections at the interfaces, a quasiparticle on
average traverses the interlayer many times. The time that a
quasiparticle effectively spends in the interlayer is proportional
to $D^{-1}$, where $D$ is the transparency of each barrier. For a
transparency of the order of $10^{-6}$, the time of flight in the
interlayer is for example of the order of $\tau  = d/Dv_F = 0.5$
ns, for a thickness of about $10$ nm and a typical Fermi-velocity
of $1.5 \times 10^6$ m/s \cite{Zehnder}.

In most double-barrier Josephson junctions Al is used as an
interlayer material and therefore the inelastic scattering time in
Al should be considered. Kaplan \textit{et al.} \cite{Kaplan}
estimated an inelastic scattering time in bulk Al of 400 ns which
is much larger than 0.5 ns. However, magnetoresistance and
microwave measurements in thin films of Al \cite{Son,Santhanam}
showed that the inelastic scattering time in thin Al films is
orders of magnitudes smaller than in bulk, namely of the order of
0.1 to 1.0 ns in films of a few to 10 nm thickness. Therefore, in
the modeling of time-dependent transport properties of
double-barrier junctions, inelastic scattering, or energy
relaxation, has to be taken into account. The inelastic scattering
comprises both electron-phonon and electron-electron scattering.

\subsection{Derivation of a microscopic model}

In this section, a microscopic model will be derived for the
quasiparticle current as function of voltage in double-barrier
Josephson junctions with low-transparent barriers. It will be
shown that the results coincide with the phenomenological model by
Heslinga and Klapwijk \cite{Heslinga}, who derived their model by
matching the population and extraction rates of the quasiparticles
in the interlayer.

In this section, the assumption will be made that
$\gamma_{{\mathop{\rm eff}\nolimits}} \gg 1$. In this
approximation, the proximity effect can be neglected, i.e.
$F^{R(A)}  = 0$. Furthermore, the spectral supercurrent
Im$I_S(E,t) = 0$,  ${\mathop{\rm Re}\nolimits} G \left( E \right)
= 1$, and $D_L  = D_T  = 1$. In this case, none of the quantities
explicitly depends on time. Then, the kinetic Eqs. (\ref{kinetic})
can be simplified to
\begin{eqnarray}
 &&\mathcal D\tau _{in} \frac{{\partial ^2 }}{{\partial x^2 }}f_L \left( {E,x} \right) - \left[ {f_L \left( {E,x} \right) - f_0 \left( E \right)} \right] = 0, \nonumber\\
 &&\mathcal D\tau _{in} \frac{{\partial ^2 }}{{\partial x^2 }}f_T \left( {E,x} \right) - f_T \left( {E,x} \right) = 0,
 \end{eqnarray}
where $f_0 \left( E \right) = \tanh (E/2T)$ and $\mathcal D$ is
the diffusion constant in the interlayer. Use is made of the fact
that $f_{T0} \left( E \right) = 0$ in equilibrium. The kinetic
equations are decoupled in this case, but $f_T$ and $f_L$ are
coupled through the boundary conditions.

\begin{figure}
\includegraphics [scale=0.8]{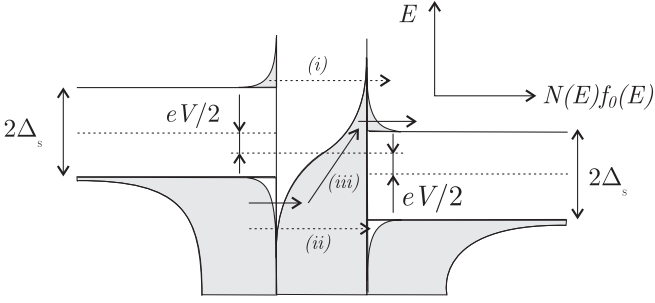}
\caption{\label{fig3_5} Semiconductor-diagram representation of
tunnel and scattering rates in a double-barrier junction at
bias-voltage $eV = \Delta_S$. The energy conserving processes,
($i$) and ($ii$), are in the case of inelastic scattering
complemented by process ($iii$).}
\end{figure}

The boundary conditions can either be obtained by simplifying the
relevant terms of the expressions that contain all harmonics, such
as Eqs. (\ref{term1}) and (\ref{term2}), or by starting from the
time-dependent boundary conditions, Eqs. (\ref{bound1}) and
(\ref{bound2}). In the latter case, the transformation to energy
space is straightforward. The right-hand sides of Eqs.
(\ref{bound1}) and (\ref{bound2}) only contain terms that depend
on time difference since $F^{R(A)} = 0$, e.g.
\begin{eqnarray}
&&{\mathop{\rm Re}\nolimits} G_1(t-t_1) \circ i\sin {\frac{{eV( {t
- t_1} )}}{\hbar}}  \circ {\mathop{\rm Re}\nolimits}
     G(t_1-t')\nonumber\\&& \circ
      f_L \left( {t_1 - t'} \right) = \int {dEe^{ - i{E\left( {t - t'} \right)}/\hbar } }
{\mathop{\rm Re}\nolimits} G \left( E \right)f_L \left( E
\right)G_-,\nonumber\\
\end{eqnarray}
where $G_-={\mathop{\rm Re}\nolimits} G_1 \left( {E + eV/2}
\right) + {\mathop{\rm Re}\nolimits} G_1 \left( {E - eV/2}
\right)$. The left-hand side of Eq. (\ref{bound1}) becomes
\begin{eqnarray}
&&D_T \gamma _B \xi \frac{\partial }{{\partial x}}f_T \left(
{t,t'} \right)\nonumber\\&& = \gamma _B \xi \frac{\partial
}{{\partial x}}\int {dEe^{ - iE\left( {t - t'} \right)} D_T \left(
E \right)f_T \left( E \right)} .
\end{eqnarray}
Hence, together with ${\mathop{\rm Re}\nolimits} G  = 1$, finally
the boundary conditions read
\begin{eqnarray}
 \gamma _B \xi  \frac{\partial }{{\partial x}}f_T \left( {E, \pm d/2} \right) =  \mp f_T \left( {E, \pm d/2} \right)N_ +\nonumber\\   - f_L \left( {E, \pm d/2} \right)N_ -   + R_ -  , \nonumber\\
 \gamma _B \xi  \frac{\partial }{{\partial x}}f_L \left( {E, \pm d/2} \right) =  \mp f_L \left( {E, \pm d/2} \right)N_ +\nonumber\\   - f_T \left( {E, \pm d/2} \right)N_ -   \pm R_ -
 ,\label{bound4}
 \end{eqnarray}
where $N_ \pm   = {\mathop{\rm Re}\nolimits} G_1 \left( {E + eV/2}
\right) \pm {\mathop{\rm Re}\nolimits} G_1 \left( {E - eV/2}
\right)$ in the superconductors and $R_ \pm   = {\mathop{\rm
Re}\nolimits} G_1 \left( {E + eV/2} \right) \times f_0 \left( {E +
eV/2} \right) \pm {\mathop{\rm Re}\nolimits} G_1 \left( {E - eV/2}
\right)f_0 \left( {E - eV/2} \right)$. The kinetic equations
provide that $ c_1  = 2a_1 D\tau _{in}  = 2a_1 \delta $ for the
Ansatz $f_T  = a_1 x^2 + b_1 x + c_1 $ and $f_L  = a_2 x^2  + b_2
x + c_2 $. Using boundary conditions (\ref{bound4}) and neglecting
terms proportional to $d^2$, the solution can be simply found. The
quasiparticle current is given by Eq. (\ref{current2}), where
$df_T /dx = b_1 $, and $b_1$ is given by
\begin{equation}
b_1  = \frac{1}{{\gamma _B \xi }}\frac{{R_ -  N_ +   - R_ + N_ - +
\left( {R_ - - f_0 N_ -  } \right)/\Gamma \tau _{in} }}{{N_ + +
1/\Gamma \tau _{in} }},
\end{equation}
where $\Gamma ^{ - 1}  = \gamma _B d\xi  /\mathcal D\hbar  = e^2
N( 0 )R_B d/\hbar $ is the tunneling injection rate into the
normal metal interlayer, $N(0)$ the unnormalized density of states
in the interlayer and $R_B$ the specific barrier resistance. Using
the fact that density of states functions are symmetric in energy
and $f_0$ is asymmetric in energy, $ ( {R_ - - f_0 N_ - } )$ can
be simplified to $2{\mathop{\rm Re}\nolimits} G_1 ( E + eV/2 )[f_0
( {E + eV/2} )-f_0 ( {E} )]$ and $R_ - N_ + - R_ + N_ - = 2
{\mathop{\rm Re}\nolimits} G_1 ( E + eV/2 ){\mathop{\rm
Re}\nolimits} G_1 ( E - eV/2 )[f_0 (E + eV/2)-f_0 ( (E - eV/2)
)]$. With $\sigma _N /2\gamma _B = e^2 N( 0 )\mathcal D/\gamma _B
= R_B^{ - 1} $, the expression for the quasiparticle current
finally becomes
\begin{eqnarray}
 &&I = \frac{2}{{eR_N }}\int\limits_{ - \infty }^\infty  {dE{\mathop{\rm Re}\nolimits} G_1 \left( {E + eV/2} \right)}  \nonumber\\
  &&\times \frac {{\mathop{\rm Re}\nolimits} G_1 \left( {E - \frac {eV}{2}} \right)F_{\_} +\left[ {f_0 \left( {E + \frac {eV}{2}} \right)-f_0 \left( {E} \right)}\right]/\Gamma \tau _{in}} {{G_++1/\Gamma \tau_{in}}
  },\nonumber\\\label{inelastic}
 \end{eqnarray}
where $F_{\_}= f_0 \left( {E + \frac{{eV}}{2}} \right) - f_0
\left( {E - \frac{{eV}}{2}} \right) $ and $G_+={\mathop{\rm
Re}\nolimits} G_1 \left( {E + eV/2} \right) + {\mathop{\rm
Re}\nolimits} G_1 \left( {E - eV/2} \right)$. This expression is
equivalent to the findings of Heslinga and Klapwijk
\cite{Heslinga}, in the limit of ${\mathop{\rm Re}\nolimits}G=1$,
who derived a model by equating the population and extraction
rates in the interlayer. Zaitsev's results for the SININ junction
in the limit of no energy relaxation \cite{Zaitsev90} coincide
with our findings as well. The equivalence of a mesoscopic or
phenomenological approach and the more rigorous Green's functions
treatment is shown by Argaman \cite{Argaman} to hold for the
equations for current. Here, we have proven that the final
expression, Eq.~(\ref{inelastic}) also follows from the Green's
function approach, using the appropriate boundary conditions.

\subsection{Influence of inelastic scattering on transport
properties}\label{section:inelastic}

\begin{figure}
\includegraphics [scale=1.25]{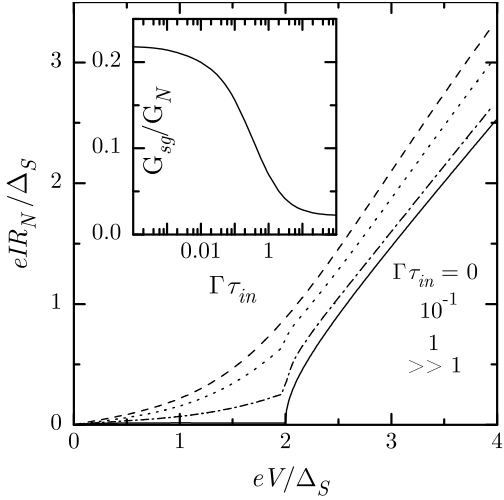}
\caption{\label{fig3_6_a} $IV$ characteristics at $k_BT/\Delta_S =
0.25$ and $\gamma_{\rm eff} \gg 1$ on the basis of Eq.
(\ref{inelastic}) for several values of the inelastic scattering
parameter $\Gamma \tau_{in}$. The inset shows the subgap
conductance at $eV = \Delta$ as function of $\Gamma \tau_{in}$. }
\end{figure}

Examples of possible tunneling processes are indicated in
Fig.~\ref{fig3_5}. In one of the processes a quasiparticle is
inelastically scattered in the interlayer. Equation
(\ref{inelastic}) coincides in the limit of strong inelastic
scattering ($\Gamma \tau _{in}  = 0$) with the known result for
two SIN tunnel junctions in series. In the absence of inelastic
scattering, Eq. (\ref{inelastic}) reduces to
\begin{eqnarray}
&&I = \int\limits_{ - \infty }^\infty  {\frac{{2dE}}{{eR_N
}}{\mathop{\rm Re}\nolimits} G_1 \left( {E + \frac{{eV}}{2}}
\right)} {{\mathop{\rm Re}\nolimits} G_1 \left( {E -
\frac{{eV}}{2}} \right) \frac {
F_-}{G_+}}.\nonumber\\\label{inelastic2}
\end{eqnarray}
Figure~\ref{fig3_6_a} shows both limiting cases as well as $IV$
curves for intermediate values of the scattering parameter, taken
in the present limit of $\gamma_{{\rm{eff}}} \gg 1$. It can be
seen in the inset of Fig.~\ref{fig3_6_a}, that inelastic
scattering enhances the subgap-conductance. This effect will be
discussed in section \ref{section:application} in order to explain
the large subgap conductance observed in double-barrier junction
measurements.

\begin{figure}
\includegraphics [scale=1.25]{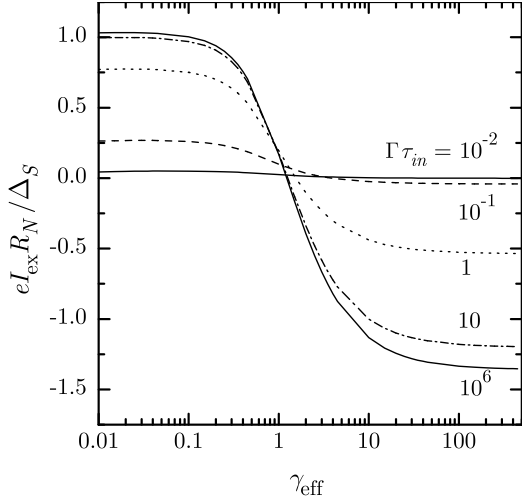}
\caption{\label{fig3_6_b} Excess and deficit current as function
of the supression parameter for several values of the inelastic
scattering parameter $\Gamma \tau_{in}$. }
\end{figure}

Equation (\ref{inelastic2}) gives a deficit current of
$eI_{def}R_N = 4\Delta_S/3$ for $eV \gg \Delta _S $, which
coincides with the findings of section \ref{section:qp}. In
analogy with the approach of section \ref{section:qp} to calculate
excess and deficit currents by summing the respective
contributions from SININ and NINIS junctions, the same can be
calculated by including inelastic scattering as well.
Figure~\ref{fig3_6_b} shows the resulting crossover from excess to
deficit current as function of the suppression parameter for
several values of the inelastic scattering parameter. For $\Gamma
\tau _{in}  < 10^{ - 1} $ only a small deficit current is
predicted at $eV \gg \Delta _S $. However, at moderate values of
$V$, i.e. $2\Delta _S  < eV < 4\Delta _S $, still a considerable
deficit current is present, as for example can be seen in
Fig.~\ref{fig3_6_a}.

\section{Application: the nature of the intrinsic shunt}\label{section:application}

The Resistively and Capacitively Shunted Junction (RCSJ) model
shows how a sinusoidal supercurrent-phase relation, a linear
quasiparticle current and a displacement-current determine the
shape of the entire \textit{IV} characteristic of a Josephson
junction. The model can also be applied to an unshunted junction,
but then the subgap resistance $R_{sg}$ appears in the expression
for the Stewart-McCumber parameter
\begin{equation}
\beta _c  = 2\pi \frac{{\left( {I_c R_N } \right)^2 C}}{{I_c \Phi
_0 }}\left( {\frac{{R_{sg} }}{{R_N }}} \right)^2 , \label{beta}
\end{equation}
where $C$ is the capacitance of the junction and $\Phi_0 (= 2.07
\times 10^{-15}$ Wb) the flux quantum. Likharev \cite{Likharev}
showed that the relation between $\beta_C$ and the presence of
hysteresis depends on the model that is used to describe the
junction (e.g. Non-linear Resistive model, with different
dependencies for the subgap conductance, and the Tunnel Junction
Microscopic model), but roughly speaking, it can be said that
$\beta_C > 1$ corresponds to hysteretic $IV$ characteristics.
Hysteresis refers here to the existence of two brances in the $IV$
curve, one going from the critical current $I_c$ to the voltage
state, and one going back at the return current ($I_R < I_c$) from
the voltage state to the state at $V=0$.

The capacitance of a double-barrier junction is not known a
priori. In Ref. ~\onlinecite{Squid} a set of Nb/Al double-barrier
Josephson junctions was fabricated in order to make SQUIDs. From
resonances in the SQUID washer, $C$ was determined to be 0.015
pF/$\mu$m$^2$, corresponding to the capacitance of two SIS
junctions in series \cite{Squid}. It is assumed that this value is
only weakly depending on the transparency of the barrier. The
dependence of $I_c$ and $R_N$ on the junction parameters, such as
$\gamma_{{\mathop{\rm eff}\nolimits}}$, follow from the modeling
of the stationary properties in Ref. ~\onlinecite{Kupriyanov2}.
The subgap conductance as function of the suppression parameter is
determined in section \ref{section:qp}.

\begin{figure}
\includegraphics [scale=1.25]{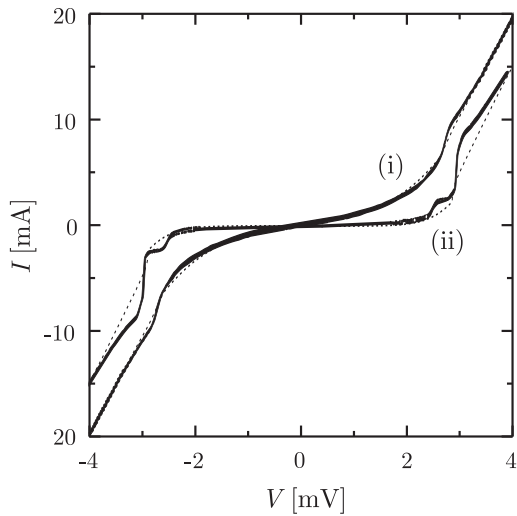}
\caption{\label{fig4_7_a} Experimental $IV$ curves (solid lines)
at 4.2 K ($i$) and 1.6 K ($ii$) together with theoretical fits
(dashed lines) with $\Gamma \tau_{in} = 0.1$ and 0.3 respectively.
}
\end{figure}

First, the regime of junctions with $\gamma_{{\mathop{\rm
eff}\nolimits}}  \gg 1$ will be discussed. For this purpose, low
critical current density Nb/Al double-barrier junctions were
fabricated according to the process of Ref. ~\onlinecite{Squid}.
Figure~\ref{fig4_7_a} shows a typical measured $IV$
characteristic, together with an $IV$ curve from the
nonequilibrium model of section \ref{section:inelastic}, where
inelastic scattering is taken into account. At 4.2 K, the
experimental and theoretical curves are very much alike, taking
into account the fact that only one free parameter was used to
fit, namely $\Gamma \tau_{in}$. From
\begin{equation}
\Gamma \tau _{in}  = \frac{{\pi T_{cS} k_B
}}{{\gamma_{{\mathop{\rm eff}\nolimits}} }}\frac{{\tau _{in}
}}{h},
\end{equation}
and the fitted $\Gamma \tau_{in} = 0.1$ and $\gamma_{{\mathop{\rm
eff}\nolimits}} = 2 \times 10^3$ (which was obtained from fitting
the critical current temperature dependence), an inelastic
scattering time $\tau_{in} = 0.3$ ns is obtained in the Al
interlayers. At 1.6 K, a magnetic field was used to suppress the
supercurrent in order to resolve the subgap quasiparticle
conductance. The deviation of the fit from the experiment around
$2\Delta_{Nb}$ is due to the nonequilibrium enhancement of the gap
in the interlayer, as described in Ref. ~\onlinecite{Capogna},
which can be included in the model by incorporating $\Delta_{Al}$.
However, the good fit well below $2\Delta_{Nb}$ allows for the
extraction of $\tau_{in} = 0.9$ ns at 1.6 K.

The values for inelastic scattering correspond to measurements by
Santanam \textit{\textit{et al.}} \cite{Santhanam} who found
$\tau_{in} = 0.2$ to $1.0$ ns in 10 nm Al films at 4.2 K, and Van
Son \textit{et al.} \cite{Son} who found $\tau_{in} = 0.8$ to
$0.9$ ns in 7 nm Al films at $T_{cAl}$. Our values of $\tau_{in} =
0.9$ ns at 1.6 K and 0.3 ns at 4.2 K indicate a scaling with
$T^{-1}$ rather than $T^{-3}$, which was found and discussed as
well by Santanam \textit{et al} \cite{Santhanam}. Note, that the
values for the inelastic scattering are much smaller than $\pi
k_BT$, which means that the stationary properties are not
influenced by $\tau_{in}$. Furthermore, from these values it is
seen that $\gamma_{{\rm eff}} \delta^{-1} \gg 1$ as long as $
\gamma_{{\rm eff}} \le 10^3$, which was used in order to obtain
Fig.~\ref{fig_Is}.

\begin{figure}
\includegraphics [scale=1.25]{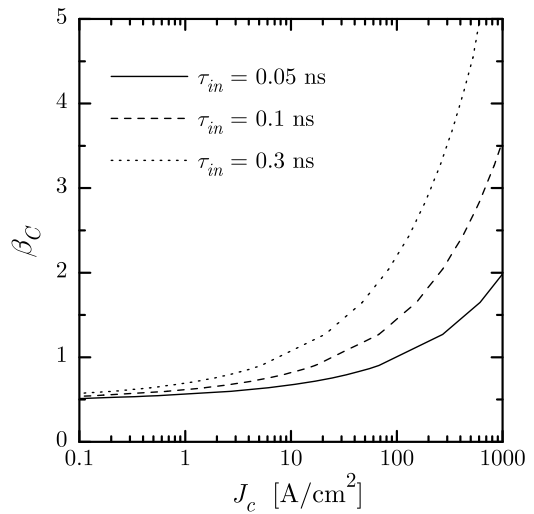}
\caption{\label{fig4_7_b} Expected $\beta_C$ as function of
critical current density from the inelastic scattering model. $T=
4.2$ K}
\end{figure}

As a measure of the subgap conductance, the theoretically expected
normalized conductance at $eV = \Delta_S$ can be found in
Fig.~\ref{fig3_6_a}, as function of the inelastic scattering
parameter $\Gamma \tau_{in}$. It can be seen that the subgap
resistance in the limit of zero inelastic scattering ($\tau _{in}
\to \infty $) is only determined by temperature. The conductance
in this limit is therefore called the thermal contribution. The
relation between subgap resistance and $\gamma_{{\mathop{\rm
eff}\nolimits}}$ is now known for a fixed value of $\tau_{in}$
since $\Gamma$ is given by $\pi k_BT/\gamma_{{\mathop{\rm
eff}\nolimits}}$.

The dependence of $I_cR_N$ on $\gamma_{{\mathop{\rm
eff}\nolimits}}$ is known from the Matsubara modeling of the
stationary properties of double-barrier junctions
\cite{Kupriyanov2}. Together with the definition of
$\gamma_{{\mathop{\rm eff}\nolimits}}$, it follows that
\begin{equation}
R_N^{ - 1}  = \frac{{e^2 k_F^2 }}{{2\pi ^2 \hbar }}\frac{{\pi k_B
T_{cS} d}}{{\hbar v_F \gamma_{{\mathop{\rm eff}\nolimits}} }},
\end{equation}
where the parameter values can be taken as $v_F = 1.5 \times 10^6$
m/s \cite{Zehnder}, $d = 6$ nm, and $T_{cNb} = 9.2$ K. Putting
these theoretical dependencies together with the experimentally
determined parameters into Eq. (\ref{beta}), provides $\beta_C$ as
function of the critical current density for junctions with
$\gamma_{{\mathop{\rm eff}\nolimits}}  \gg 1$, see
Fig.~\ref{fig4_7_b}.

The shunting behavior can physically be explained as follows. A
direct transfer process of quasiparticles from one electrode to
the other is prohibited when the quasiparticle energy falls within
the gap of the other electrode. However, by scattering
inelastically in the interlayer, the quasiparticles are
redistributed over energy, allowing some quasiparticles to enter
the other electrode, which results in an enhanced conductance, as
illustrated in Fig.~\ref{fig3_5}. The amount of quasiparticles
that get scattered inelastically increases for decreasing barrier
transparencies, since the effective lifetime of a quasiparticle in
the interlayer is then increased. For strong inelastic scattering,
the double-barrier junction can be regarded as a series connection
of an SIN and NIS junction, where the energy distibution function
in the interlayer is the equilibrium Fermi function $f_0 =$
tanh$(E/2k_BT)$. Here it should be note that the assumption is
made that the inelastic scattering is dominated by electron-phonon
interactions, and that there is coupling between the interlayer
and a heat bath. It is known \cite{Pierre}, that in the contrary
non-adiabatic limit of MAR and strong electron-electron
interactions,  the energy distribution in the interlayer is given
by the Fermi function at temperature $k_B T = \Delta + eV$.

\begin{figure}
\includegraphics [scale=1.25]{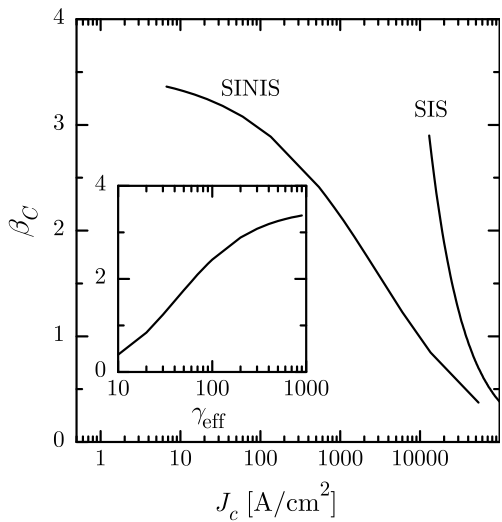}
\caption{\label{fig4_8_a} Theoretical model for $\beta_C$ as
function of $J_c$ (and as function of $\gamma_{{\mathop{\rm
eff}\nolimits}}$ in the inset), based on the contribution of
Andreev channels to the subgap conductance in high-$J_c$ junctions
at $T = 4.2$ K, in comparison with the hysteresis of SIS
junctions.}
\end{figure}

\begin{figure}
\includegraphics [scale=1.25]{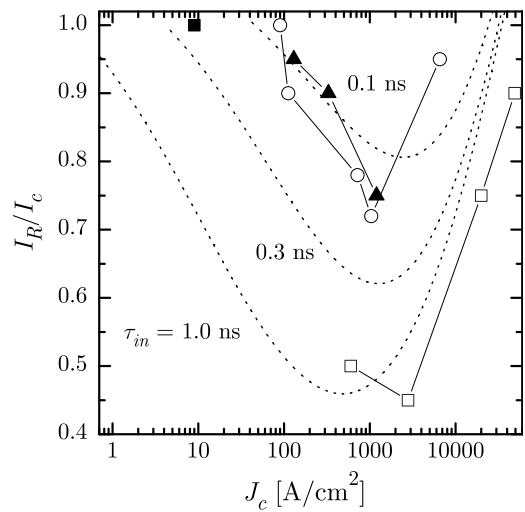}
\caption{\label{fig4_8_b} Theoretical model (dotted line) for the
ratio of return and critical current, based on the sum of the
shunting contributions at 4.2 K from both inelastic scattering and
Andreev channels, for several inelastic scattering times.
Experimental data are shown from this paper ($ \blacksquare $),
Ref. ~\onlinecite{Dima} ($\bigcirc$), Ref. ~\onlinecite{Sugiyama}
($\square$) and Ref. ~\onlinecite{Maezawa} ($\blacktriangle$).}
\end{figure}

In order to understand the intrinsic shunt of all Al-based
double-barrier junctions, the regime of high-$J_c$ junctions
(typically larger than 100 A/cm$^2$) should be considered as well.
The second contribution to the subgap conductance is due to the
Andreev reflection processes at the two superconductor-normal
metal interfaces, which was formally introduced by the term
Re$(F)$Re$(F_S)$. The Andreev channels open at high transparency
of the interface barriers. In first order, this contribution is
independent of temperature, but it depends on the suppression
parameter, which is shown in the inset of Fig.~\ref{fig3_4_a} for
a fixed temperature. For the practical range of parameters, this
means that the contribution is inversely proportional to
$\gamma_{{\mathop{\rm eff}\nolimits}}$. Figure~\ref{fig4_8_a}
shows the resulting hysteresis as function of critical current
density. Figure~\ref{fig4_8_a} predicts that non-hysteretic
double-barrier junctions can be obtained with critical current
densities of the order of 10 kA/cm$^2$ and higher. In order to
make the comparison with SIS junctions, a similar curve has been
calculated based on Eq. (\ref{beta}) and plotted in the same
Fig.~\ref{fig4_8_b} In this calculation it was assumed that $C =
3.0 \mu$F/cm$^2$, $I_cR_N = 2.0$ mV and $R_{sg} = 2R_N$. A bigger
subgap resistance will shift the SIS curve even more to the right.

Summing up all contributions to the subgap conductance provides
the theoretical curve in Fig.~\ref{fig4_8_b} for several values of
$\tau_{in}$ and $d = 6$ nm, where Zappe's equation \cite{Zappe}
was used to calculate the ratio of return and critical current
$I_R/I_c$ from $\beta_C$. The summation is performed in a
straightforward way, since the contributions due to inelastic
scattering and the opening of Andreev channels do not overlap,
i.e. they occur in separate regimes of the suppression parameter.

An increase in $\tau_{in}$ is seen to increase the hysteresis and
shift the maximum hysteresis to lower values of $J_c$. A thicker
interlayer will both decrease $J_c$, as well as shift the curve
upward since $\Gamma \tau_{in}$ is larger in this case. A decrease
in temperature rapidly enhances the hysteresis, since both the
thermal contribution to the subgap conductance decreases as well
as the contribution of inelastic scattering, since $\tau_{\rm in}$ increases
with temperature. This explains the strong influence of
temperature on hysteresis as observed in experiments, which is
stronger than could be expected from an increase in $I_c$ alone.

Observed experimental $I_R/I_c$ values are shown as well in
Fig.~\ref{fig4_8_b} and it can be concluded that the experiments
are now qualitatively and quantitatively very well explained by
the model in the sense that both the non-monotonic hysteresis
dependence on critical current density as well as the actual
hysteresis values are obtained.

\section{Conclusion}

Time-dependent and nonequilibrium transport properties of SINIS
junctions have been studied by means of a microscopic Green's
function approach.

The kinetic equations for the longitudinal and transverse energy
distribution functions are derived from the Keldysh-Usadel
equation. The appropriate boundary conditions are derived by
starting from the Kupriyanov-Lukichev boundary conditions and by
applying Gauge transformations in the electrodes. The resulting
set of equations has a recurrent nature, in terms of coupling of
Green's functions to higher harmonics as well as to functions with
shifted energy arguments. This lays out a theoretical framework to
microscopically study time-dependent problems in superconducting -
normal metal devices.

We apply this formalism in order to develop a theory of the subgap
conductance of SINIS junctions. This conductance is a very
favorable feature for applications but so far not understood on a
microscopic level. In the adiabatic limit of a small voltage and a
large suppression parameter, the time-dependencies simplify and
the equations are solved to determine the dissipative current in
double-barrier Josephson junctions. Known limiting cases, such as
the SININ' junction, are reproduced. Excess and deficit current
are determined as function of the suppression parameter and the
asymmetry between the barriers. Excess current as high as $eI_{ex}
R_N  \simeq 1.05\Delta _S $ can exist in double-barrier junctions
in the symmetric case for $\gamma_{{\mathop{\rm eff}\nolimits}}
\ll 1$, and maximum deficit current is reached in the symmetric
case for $\gamma_{{\mathop{\rm eff}\nolimits}}  \gg 1$. The subgap
conductance enhancement by decreasing $\gamma_{{\mathop{\rm
eff}\nolimits}}$ is caused by the opening of Andreev channels.

It is found that the time-dependent nonequilibrium contribution to
the energy distribution function gives rise to a nonzero averaged
supercurrent in the presence of a voltage bias. This effect should
be observable in double-barrier junction experiments.

In contrast to most studied mesoscopic systems, inelastic
scattering in the interlayer of double-barrier junctions can have
a strong influence on the electronic transport even in very short
devices. A microscopic derivation of the dependence of the
transport properties on the inelastic scattering parameter is
given. $IV$ characteristics show an enhanced subgap conductance
for increased inelastic scattering rates.

The actual value for the inelastic scattering time in the
interlayer of experimentally realized devices was obtained by
fitting the microscopic model. The inelastic scattering values
explain, together with the opening of Andreev channels, the nature
of the intrinsic shunt in double-barrier junctions.

\begin{acknowledgments}
This project was supported by the Dutch Foundation for Fundamental
Research on Matter (FOM). We thank H. Hilgenkamp, T.M. Klapwijk,
Th. Sch\"apers, V.S. Shumeiko, and A.D. Zaikin for valuable
discussions. FKW is grateful to B.\ Pannetier and H.\ Courtois for
hospitality and discussions during the very early stages of this
project.
\end{acknowledgments}

\appendix

\section{Derivation of the boundary conditions}\label{appendix:bound}

From the definition of the Green's functions in Keldysh $\times$
Nambu space, Eq. (\ref{Keldysh}), a boundary condition can be
written for each of the matrix elements of Eq. (\ref{KLbound})
\begin{eqnarray}
 && \gamma _B \xi \hat G^R \frac{d}{{dx}}\hat G^R  = \hat G^R \hat G_1^R  - \hat G_1^R \hat G^R , \nonumber\\
 && \gamma _B \xi \hat G^A \frac{d}{{dx}}\hat G^A  = \hat G^A \hat G_1^A  - \hat G_1^A \hat G^A , \nonumber\\
 && \gamma _B \xi \left( {\hat G^R \frac{d}{{dx}}\hat G^K  + \hat G^K \frac{d}{{dx}}\hat G^A } \right) \nonumber\\
 &&  = \hat G^R \hat G_1^K  + \hat G^K \hat G_1^A  - \hat G_1^R \hat G^K  - \hat G_1^K \hat G^A
 .\label{A1}
\end{eqnarray}
With definition (\ref{param}), the left-hand side of the latter of
these three boundary conditions becomes
\begin{eqnarray}
 &&\gamma _B \xi \left[ {\left( {\hat G^R \frac{d}{{dx}}\hat G^R } \right)\hat f  + \hat G^R \hat G^R \frac{d}{{dx}}\hat f } \right. \nonumber\\
 &&\left. { - \hat G^R \frac{d}{{dx}}\left( {\hat f \hat G^A } \right) + \hat G^R \hat f \frac{d}{{dx}}\hat G^A  + \hat f \hat G^A \frac{d}{{dx}}\hat G^A } \right].\nonumber\\
\end{eqnarray}
With the aid of the first two Eqs. in (\ref{A1}) this can be
rewritten into
\begin{eqnarray}
 &&\gamma _B \xi \left[ {\hat G^R \hat G^R \frac{d}{{dx}}\hat f  - \hat G^R \frac{d}{{dx}}\left( {\hat f \hat G^A } \right)} \right.\left. { + \hat G^R \hat f \frac{d}{{dx}}\hat G^A } \right] \nonumber\\
 && + \left( {\hat G^R \hat G_1^R  - \hat G_1^R \hat G^R } \right)\hat f  - \hat f \left( {\hat G^A \hat G_1^A  - \hat G_1^A \hat G^A } \right).
\end{eqnarray}
By making use of the normalization condition $G^{R(A)} G^{R(A)}=1$
and the definition for $\hat f$ and $\hat f_1 $, Eq.
(\ref{notation}), this can be futher rewritten as
\begin{eqnarray}
 &&\gamma _B \xi \left[ {\frac{d}{{dx}}\left( {f_{L}  + \hat \tau _3 f_{T} } \right) - \hat G^R \frac{d}{{dx}}\left( {f_{L}  + \hat \tau _3 f_{T} } \right)\hat G^A } \right] \nonumber\\
 && + \left( {\hat G^R \hat G_1^R  - \hat G_1^R \hat G^R } \right)\left( {f_{L}  + \hat \tau _3 f_{T} } \right) \nonumber\\
 && - \left( {f_{L}  + \hat \tau _3 f_{T} } \right)\left( {\hat G^A \hat G_1^A  - \hat G_1^A \hat G^A } \right),
\end{eqnarray}
which is equal to
\begin{eqnarray}
 &&\gamma _B \xi \left[ {\left( {1 - \hat G^R \hat G^A } \right)\frac{d}{{dx}}f_{L}  + \left( {\hat \tau _3  - \hat G^R \hat \tau _3 \hat G^A } \right)\frac{d}{{dx}}f_{T} } \right] \nonumber\\
 && + \left( {\hat G^R \hat G_1^R  - \hat G_1^R \hat G^R  - \hat G^A \hat G_1^A  + \hat G_1^A \hat G^A } \right)f_{L}  \nonumber\\
 && + \left[ {\left( {\hat G^R \hat G_1^R  - \hat G_1^R \hat G^R } \right)\hat \tau _3  - \hat \tau _3 \left( {\hat G^A \hat G_1^A  - \hat G_1^A \hat G^A } \right)} \right]f_{T}
 .\nonumber\\\label{A5}
\end{eqnarray}
The right-hand side of the last boundary condition in Eq.
(\ref{A1}) can be rewritten with the aid of Eq. (\ref{param}) into
\begin{eqnarray}
 &&\hat G^R \hat G_1^R \hat f_1  - \hat G_1^R \hat f_1 \hat G^A  - \hat G^R \hat f_1 \hat G_1^A  + \hat f_1 \hat G_1^A \hat G^A  \nonumber\\
 && - \hat f \hat G^A \hat G_1^A  - \hat G_1^R \hat G^R \hat f  + \hat G^R \hat f \hat G_1^A  + \hat G_1^R \hat f \hat G^A .\nonumber\\
\end{eqnarray}
With definition Eq. (\ref{notation}) for $\hat f$ and $\hat f_1 $
this becomes
\begin{eqnarray}
 &&\left[ {\hat G^R \left( {\hat G_1^R  - \hat G_1^A } \right) - \left( {\hat G_1^R  - \hat G_1^A } \right)\hat G^A } \right]f_{L1}  \nonumber\\
 && + \left[ {\hat G^R \left( {\hat G_1^R \hat \tau _3  - \hat \tau _3 \hat G_1^A } \right) - \left( {\hat G_1^R \hat \tau _3  - \hat \tau _3 \hat G_1^A } \right)\hat G^A } \right]f_{T1}  \nonumber\\
 &&\left[ {\left( {\hat G^R  - \hat G^A } \right)\hat G_1^A  - \hat G_1^R \left( {\hat G^R  - \hat G^A } \right)} \right]f_{L} \nonumber \\
 && + \left[ {\hat G^R \hat \tau _3 \hat G_1^A  - \hat \tau _3 \hat G^A \hat G_1^A  + \hat G_1^R \hat \tau _3 \hat G^A  - \hat G_1^R \hat G^R \hat \tau _3 } \right]f_{T}
 .\nonumber\\\label{A7}
\end{eqnarray}
Equating left-hand and right-hand side of the last boundary
condition in Eq. (\ref{A1}), i.e. Eqs. (\ref{A5}) and (\ref{A7})
respectively, finally gives the form of the boundary condition as
presented in Eq. (\ref{boundstart}).

\section{Time convolutions in energy space}\label{appendix:convolutions}

The expression for a convolution of two functions,
\begin{equation}
a \circ b\left( {t,t'} \right) = \int\limits_{ - \infty }^\infty
{dt_1 a\left( {t,t_1 } \right)} ^{\rm{ }} b\left( {t_1 ,t'}
\right),
\end{equation}
can be transformed by changing variables
\begin{eqnarray}
&&a \circ b\left( {t,t'} \right) =\nonumber\\ &&\int\limits_{ -
\infty }^\infty {dt_1 a\left( {t - t_1 ,\frac{{t - t_1 }}{2}}
\right)} ^{\rm{ }} b\left( {t_1  - t',\frac{{t_1  - t'}}{2}}
\right).
\end{eqnarray}
Subsequently, a Fourier transform to energy-frequency space can be
made
\begin{eqnarray}
  && a \circ b\left( {t,t'} \right) = \int\limits_{ - \infty }^\infty  {dt_1 d\omega dEd\omega 'dE'a\left( {E,\omega } \right)e^{iE\left( {t - t_1 } \right)/\hbar}}\nonumber\\
  && e^{i\omega \left( {t + t_1 } \right)/2\hbar }  b\left( {E',\omega '} \right)e^{iE'\left( {t_1  - t'} \right)/\hbar } e^{\frac{{i\omega '\left( {t_1  + t'} \right)}}{{2\hbar}}}\nonumber\\
  && = \sum\limits_{n'n'} {\int\limits_{ - \infty }^\infty  {dt_1 dEdE'a_n \left( E \right)e^{iE\left( {t - t_1 } \right)/\hbar } e^{in\omega _0 \left( {t + t_1 } \right)/2\hbar }}}\nonumber\\
  && b_{n'} \left( {E'} \right)} e^{iE'\left( {t_1  - t'} \right)/\hbar } e^{in'\omega _0 \left( {t_1  + t'} \right)/2\hbar\nonumber\\
  && = \sum\limits_{n'n'} {\int\limits_{ - \infty }^\infty  {dEdE'a_n \left( E \right)e^{iEt/\hbar } e^{in\omega _0 t/2\hbar } } b_{n'} \left( {E'} \right)} ^{\rm{ }}\nonumber\\
  && e^{ - iE't'/\hbar } e^{in'\omega _0 t'/2\hbar }\delta \left( { - E + n\omega _0 /2 + E' + n'\omega _0 /2} \right)\nonumber\\
  && = \sum\limits_{n'n'} {\int\limits_{ - \infty }^\infty  {dE'a_n \left( {E' + \frac{{n + n'}}{2}\omega _0 } \right)} ^{\rm{ }} b_{n'} \left( {E'} \right)} ^{\rm{ }}\nonumber\\
  && e^{ - iE'\left( {t - t'} \right)/\hbar } e^{i\frac{n}{2}\omega _0 \left( {t - t'} \right)/\hbar } e^{i\frac{{n' + n}}{{2\hbar }}\omega _0 \left( {t + t'} \right)}.
\end{eqnarray}
With an energy-shift $E = \tilde E + n\omega _0 /2$ this becomes
\begin{eqnarray}
a \circ b\left( {t,t'} \right) &=& \sum\limits_{n'n'}
{\int\limits_{ - \infty }^\infty  {dEa_n \left( {E' +
\frac{{n'\omega _0 }}{2}} \right)} ^{\rm{ }} b_{n'} \left( {E -
\frac{{n\omega _0 }}{2}} \right)^{\rm{ }} }\nonumber\\
&& \times e^{ - i\frac{{E\left( {t - t'} \right)}}{\hbar }}
e^{i\frac{{n' + n}}{{2\hbar }}\omega _0 \left( {t + t'} \right)} .
\end{eqnarray}
The triple products in the boundary conditions (3.34) and (3.35)
can be worked out in the same manner. The sine and cosine terms
cause shifts in the arguments. An example of the result of a
triple convolution is given in Eq. (\ref{term2}).


\begin{references}

\bibitem{Larkin} Nonequilibrium superconductivity, edited by D.N. Langenberg and
A.I. Larkin, Elsevier, Amsterdam, 1986.

\bibitem{Gray}Nonequilibrium superconductivity, edited by K.E. Gray, Plenum,
New-York, 1981.

\bibitem{Kopnin}Theory of nonequilibrium superconductivity, N.B.
Kopnin, Clarendon Press, Oxford, 2001.

\bibitem{Tinkham2}M. Tinkham and J. Clarke, Phys. Rev. Lett. \textbf{28}, 1366 (1972).

\bibitem{Clarke}J. Clarke in Nonequilibrium superconductivity, edited by D.N.
Langenberg and A.I. Larkin, Elsevier, Amsterdam, 1986.

\bibitem{Eliashberg}G.M. Eliashberg and B.I. Ivlev in Nonequilibrium
superconductivity, edited by D.N. Langenberg and A.I. Larkin,
Elsevier, Amsterdam, 1986.

\bibitem{Volkov}A.F. Volkov, Phys. Rev. Lett. \textbf{74}, 4730 (1995).

\bibitem{Wilhelm}F.K. Wilhelm, G. Sch\"on, and A.D. Zaikin, Phys. Rev. Lett. \textbf{81}, 1682
(1998).

\bibitem{Yip}S. K. Yip, Phys. Rev. B \textbf{58}, 5803 (1988).

\bibitem{Morpurgo}A.F. Morpurgo, T.M. Klapwijk, and B.J. van Wees, Appl. Phys. Lett.
\textbf{72}, 966 (1998).

\bibitem{Baselmans}J.J.A. Baselmans, A.F. Morpurgo, B.J. van Wees, and T.M. Klapwijk,
Nature \textbf{397}, 43 (1999).

\bibitem{Wees}B.J. van Wees, K.-M.H. Lenssen, and C.J.P.M. Harmans, Phys. Rev. B
\textbf{44}, 470 (1991).

\bibitem{Samuelsson}P. Samuelsson, V.S. Shumeiko, and G. Wendin, Phys. Rev. B \textbf{56}, 5763
(1997).

\bibitem{Ilhan2}H.T. Ilhan, H.V. Demir and P.F. Bagwell, Phys. Rev. B \textbf{58}, 15120
(1998).

\bibitem{Thomas}Th. Sch\"apers, J. Malindretos, K. Neurohr, S. Lachenmann, A. van der Hart, G. Crecelius, H. Hardtdegen, H.
L\"uth, and A.A. Golubov, Appl. Phys. Lett. \textbf{73}, 2348
(1998).

\bibitem{Richter}A. Richter, Adv. Solid St. Phys. \textbf{40}, 321 (2000).

\bibitem{Samuelsson2}P. Samuelsson, J. Lantz, V.S. Shumeiko, and G. Wendin, Phys. Rev.
B \textbf{62}, 1319 (2000).

\bibitem{Pierre} F. Pierre, A. Anthore, H. Pothier, C. Urbina, and
D. Esteve, Phys. Rev. Lett. \textbf{86}, 1078 (2001).

\bibitem{Bardas}D. Averin and A. Bardas, Phys. Rev. Lett. \textbf{75},
1831 (1995); A. Bardas and D.V. Averin, Phys. Rev. B \textbf{56},
8518 (1997).

\bibitem{Cuevas}J.C. Cuevas, A. Mart\'in-Rodero, and A. Levy Yeyati,
Phys. Rev. B \textbf{54}, 7366 (1996).

\bibitem{Bezuglyi2}E.V. Bezuglyi, E.N. Bratus, V.S. Shumeiko, G.
Wendin, and H. Takayanagi, Phys. Rev. B \textbf{62}, 14439 (2000).

\bibitem{Brinkman}A. Brinkman and A.A. Golubov, Phys. Rev. B \textbf{61}, 11297 (2000).

\bibitem{Schep}K.M. Schep and G.E.W. Bauer, Phys. Rev. Lett. \textbf{78},
3015 (1997).

\bibitem{Naveh}Y. Naveh, V. Patel, D.V. Averin, K.K. Likharev, and
J.E. Lukens, Phys. Rev. Lett. \textbf{85}, 5404 (2000).

\bibitem{RSFQ}K.K. Likharev and V.K. Semenov, IEEE Trans. Appl.
Supercond. \textbf{1}, 3 (1991).

\bibitem{Kupriyanov2}M.Yu. Kupriyanov, A. Brinkman, A.A. Golubov, M. Siegel, and H. Rogalla, Physica C \textbf{326-327}, 16 (1999).

\bibitem{Kupriyanov3}D. Balashov, F.-Im. Buchholz, H. Schulze,
M.I. Khabipov, R. Dolata, M.Yu. Kupriyanov, and J. Niemeyer,
Supercond. Sci. Technol. \textbf{13}, 244 (2000).

\bibitem{Zaitsev90}A.V. Zaitsev, Pis'ma Zh. Eksp. Teor. Fiz. \textbf{51}, 35 (1990) [Sov. Phys.
JETP Lett. \textbf{51}, 41 (1990)]; A.V. Zaitsev, Pis'ma Zh. Eksp.
Teor. Fiz. \textbf{55}, 66 (1992) [Sov. Phys. JETP Lett.
\textbf{55}, 67 (1992)].

\bibitem{Volkov2}A.F. Volkov, A.V. Zaitsev, and T.M. Klapwijk, Physica C \textbf{210},
21 (1993).

\bibitem{Zaitsev99}A.V. Zaitsev, A.F. Volkov, S.W.D. Bailey, and
C.J. Lambert, Phys. Rev. B 60, 3559 (1999).

\bibitem{Lempitskii}S.V. Lempitskii, Zh. Eksp. Teor. Fiz. \textbf{85}, 1072 (1983) [Sov. Phys.
JETP \textbf{58}, 624 (1983)].

\bibitem{Kadin}A.M. Kadin, Supercond. Sci. Technol. \textbf{14}, 276 (2001).

\bibitem{Volkov3}A.F. Volkov and T.M. Klapwijk, Phys. Lett. A \textbf{168}, 217 (1992).

\bibitem{Keldysh}L.V. Keldysh, Zh. Eksp. Teor. Fiz. \textbf{47}, 1515 (1964) [Sov. Phys.
JETP \textbf{20}, 1018 (1965)].

\bibitem{Rammer}J. Rammer and H. Smith, Rev. Mod. Phys. \textbf{58}, 323 (1986).

\bibitem{Schmid}A. Schmid in Nonequilibrium superconductivity, edited by K.E.
Gray, Plenum, New-York, 1981.

\bibitem{Schon}G. Sch\"on in Nonequilibrium Superconductivity, edited by D.N.
Langenberg and A.I. Larkin, Elsevier, Amsterdam, 1986.

\bibitem{Larkin2}A.I. Larkin and Yu.N. Ovchinnikov in Nonequilibrium
Superconductivity, edited by D.N. Langenberg and A.I. Larkin,
Elsevier, Amsterdam, 1986.

\bibitem{Belzig}W. Belzig, F.K. Wilhelm, C. Bruder, G. Sch\"on, and A.D. Zaikin,
Superlatt. Microstruc. \textbf{25}, 1251 (1999).

\bibitem{Larkin3}A.I. Larkin and Yu.N. Ovchinnikov, Zh. Eksp. Teor. Fiz. \textbf{73}, 299
(1977) [Sov. Phys. JETP \textbf{46}, 155 (1977)].

\bibitem{Zaitsev84}A.V. Zaitsev, Zh. Eksp. Teor. Fiz. \textbf{86}, 1742 (1984) [Sov. Phys.
JETP \textbf{59}, 1015 (1984)].

\bibitem{Kupriyanov}M.Yu. Kupriyanov and V.F. Lukichev, Zh. Eksp. Teor. Fiz. \textbf{94}, 139
(1988) [Sov. Phys. JETP \textbf{67}, 1163 (1988)].

\bibitem{Galaktionov}A.V. Galaktionov and A.D. Zaikin, Phys. Rev.
B \textbf{65}, 184507 (2002).

\bibitem{Charlat} F.\ Zhou, P.\ Charlat, B.\ Spivak, and B.\ Pannetier,
J.\ Low Temp.\ Phys.\ {\bf 110}, 841 (1998).

\bibitem{Bezuglyi}E.V. Bezuglyi, V.S. Shumeiko, and G. Wendin,
cond-mat/0303432 (2003).

\bibitem{Heikkila}T.T. Heikkil\"a, J. S\"arkk\"a, and F.K. Wilhelm,
Phys. Rev. B \textbf{66}, 184513 (2002).

\bibitem{Thomas2}Th. Sch\"apers, V.A. Guzenko, R.P. M\"uller, A.A. Golubov, A. Brinkman,
G. Crecelius, A. Kaluza, and H. L\"uth, Phys. Rev. B. \textbf{67},
014522 (2003).

\bibitem{Schmid2}A. Schmid and G. Sch\"on, J. Low Temp. Phys. \textbf{20}, 207 (1975).

\bibitem{Peltier} T.T.\ Heikkil\"a, T.\ V\"ansk\"a, and F.K.\ Wilhelm,
Phys.\ Rev.\ B {\bf 67}, 100502 (2003).

\bibitem{Argaman}N. Argaman, Superlatt. Microstruc. \textbf{25}, 861 (1999).

\bibitem{Dubos} P.\ Dubos, H.\ Courtois, O.\ Buisson, and B.\ Pannetier,
Phys.\ Rev.\ Lett.\ {\bf 87}, 206801 (2001).

\bibitem{Zehnder}A. Zehnder, Ph. Lerch, S.P. Zhao, Th. Nussbaumer, E.C.
Kirk, and H.R. Ott, Phys. Rev. B \textbf{59}, 8875 (1999).

\bibitem{Kaplan}S.B. Kaplan, C.C. Chi, D.N. Langenberg, J.J. Chang, S. Jafarey,
and D.J. Scalapino, Phys. Rev. B \textbf{14}, 4854 (1976).

\bibitem{Son}P.C. van Son, J. Romijn, T.M. Klapwijk, and J.E. Mooij, Phys. Rev.
B \textbf{29}, 1503 (1984).

\bibitem{Santhanam}P. Santhanam and D.E. Prober, Phys. Rev. B \textbf{29}, 3733 (1984).

\bibitem{Heslinga}D.R. Heslinga and T.M. Klapwijk, Phys. Rev. B \textbf{47}, 5157 (1993).

\bibitem{Likharev}K.K. Likharev, Dynamics of Josephson Junctions and Circuits, Taylor $\&$ Francis, 1992.

\bibitem{Squid}E. Bartolom\'e, A. Brinkman, J. Flokstra, A.A. Golubov, and H. Rogalla, Physica C \textbf{340}, 93 (2000).

\bibitem{Capogna}L. Capogna, G. Burnell, and M. Blamire, IEEE Trans. Appl. Supercond. \textbf{7}, 2415 (1997).

\bibitem{Zappe}H.H. Zappe, J. Appl. Phys. \textbf{44}, 1371 (1973).

\bibitem{Dima}D. Balashov, F.-Im. Buchholz, H. Schulze, M.I. Khabipov, R. Dolata, M.Yu. Kupriyanov, and J. Niemeyer, Supercond. Sci. Technol. \textbf{13}, 244 (2000).

\bibitem{Sugiyama}H. Sugiyama, A. Yanada, M. Ota, A. Fujimaki, and H. Hayakawa, Jpn. J. Appl. Phys. \textbf{36}, L1157 (1997).

\bibitem{Maezawa}M. Maezawa and A. Shoji, Appl. Phys. Lett. \textbf{70}, 3603 (1997).


\end{references}
\end{document}